\documentclass[%
 reprint, amsmath,amssymb, aps, superscriptaddress,times]{revtex4-1}
\usepackage{subfig}
\usepackage{graphicx}
\usepackage{physics}
\begin{document}
	
	
	\title{Dynamics of observables in a $q$-deformed harmonic oscillator}
	
	\author{Aditi Pradeep}
	\affiliation{%
		Department of Physics, National Institute of Technology Calicut, Kozhikode 673601, India
	}%
	
	\author{S. Anupama}%

	\author{C. Sudheesh}
	\email{sudheesh@iist.ac.in}
	\affiliation{
		Department of Physics, Indian Institute of Space Science and Technology, Thiruvananthapuram 695547, India\\
	}%
	\date{\today}%
	
	\begin{abstract}
		Chaos in classical systems has been studied in plenty over many years. Although the search for chaos in quantum systems has been an area of prominent research over the last few decades, the detailed analysis of many inherently chaotic quantum systems based on expectation values of dynamical variables has not been reported in the literature. In this paper, we extend the study of dynamical behaviour using expectation values of variables to a $q$-deformed harmonic oscillator. The system is found to be periodic, quasi-periodic or chaotic depending on the values of the deformation parameter $q$ and the deformed coherent amplitude $\alpha_{q}$, thus enabling us to explicitly classify the chaotic nature of the system on the basis of these parameters. The chaotic properties of the system are clearly illustrated through recurrence plots, power spectra, first-return-time distributions and Lyapunov exponents of the time series obtained for the expectation values of the dynamic variables.   
	\end{abstract}
\maketitle
\section{Introduction}
Chaos is a generic term used to describe systems that have a strong sensitivity towards initial conditions. In classical chaotic systems such as multiple coupled anharmonic oscillators, asteroid orbits, double rod pendulums and so on, a small perturbation in the initial conditions manifests itself as an exponential change in the system trajectories over time. Although several chaotic dynamical systems have been studied in the classical regime, the search for its counterpart in the quantum regime posed some consequences.

Bohr's correspondence principle, which states that the behaviour of quantum systems reproduces their classical counterparts in the limit of large quantum numbers, mandates the appearance of chaos in quantum systems. However, the  discreteness of the quantum energy levels and the solutions to the Schrödinger equation, restricts them to a quasi-periodic behaviour \cite{Jensen}. As a result, a common approach has been to quantize existing classical systems and analyse how these systems behave in a quantised scenario \cite{Casati_Springer,Casati}. These systems were reported to show chaotic behaviour equivalent to their classical counterparts only for short periods of time. Another relatively popular methodology in this direction has been to identify chaotic analogues from quantum systems in the semi-classical limit. However, these systems showed chaos only under limiting conditions \cite{Vergini,Wilkinson}. On the other hand, some authors used the expectation values of dynamical variables to study the chaotic behaviour of quantum systems \cite{Sudheesh, Sudheesh2, Athreya}. Here, we advance this analysis by extending it to $q$-deformed systems. The deformation of the Lie algebra engenders unusual properties in these deformed quantum systems, thus deeming them to be potential candidates to search for chaotic behaviour in quantum mechanics.   

The $q$-deformed oscillators have applications spanning diverse fields in physics. Recently, $q$-deformation had been applied in cosmology to study the cosmic microwave background radiation \cite{Zeng}. $q$-deformed algebras have also been used to construct observables with cosmological constants in loop quantum gravity \cite{Dupuis}. $q$-deformed bosons have been explored as tools for realization of quasibosons and modelling quasiparticles \cite{Gavrilik}. These quasibosons and quasiparticles have potential applications in subnuclear physics and quantum information theory. An important application of $q$-deformed algebras in quantum information theory is to construct quantum logic gates, for e.g., using $q$-deformed harmonic oscillator algebras \cite{Altinas}. Digital watermarking using one-dimensional ergodic chaotic maps with Tsallis type of $q$-deformation has been developed recently \cite{Behnia}. In nanoscience, low temperature behaviour of deformed fermion gas models have been used to study the interactions of quasiparticles in nanomaterials \cite{Algin}. Moreover, $q$-deformed bosonic exciton gas has been shown to constitute the high density limit of Frenkel excitons providing valuable insights into their properties within nanomaterials \cite{Zeng_cheng}. In condensed matter physics, $q$-deformed Einstein's model has been used to describe the specific heat of solids by introducing temperature fluctuations through the parameter $q$ \cite{Guha}. Additionally, the class of Fibonacci oscillators which use deformed algebras have been applied to study the thermodynamics of crystalline solids using the Debye model \cite{Marinho}  and to investigate $q$-deformed diamagnetization \cite{Marinho_Pha}. Further, physical realisation of the $q$-deformed harmonic oscillator has been demonstrated using an RLC circuit \cite{Batouli}.

In the present paper, we report the study of the dynamical behaviour of math-type $q$-deformed harmonic oscillator system which is widely studied in quantum optics \cite{Dey,Sivakumar,Recamier,Jayakrishnan}. The deformation of the Lie algebra causes the energy eigenvalues of the $q$-deformed harmonic oscillator to deviate from the linearly increasing energy eigenvalues of the non-deformed harmonic oscillator. This hints at the possibility of chaotic behaviour in this system, which we will investigate in this study. This paper is structured as follows. Section \ref{sec2} is divided into three subsections. In section \ref{sub1}, we introduce the basic math type $q$-deformed harmonic oscillator. We talk about the time evolution of the $q$-deformed variables in section \ref{sub2} and in section \ref{sub3} we describe the various tools that were used in this analysis. In section \ref{sec3}, we present our results in five subsections. Finally, we conclude in section \ref{sec4} by briefly summarising our results and observations.   
\section{Time evolution with $q$-deformation}
\label{sec2}
\subsection{The $q$-deformed Harmonic Oscillator} 
\label{sub1}
The pure quantum harmonic oscillator can be described by the Hamiltonian $H=({1}/{2})(aa^{\dagger}+a^{\dagger}a)\hbar\omega$, where $a, a^{\dagger}$ correspond to the non-deformed annihilation and creation operators, respectively. In this study, we work with the units $\hbar=\omega=1$. It is to be noted that the scaling of these terms does not affect the results of this study in anyway. The above Hamiltonian produces an eigenvalue, $E_{n}= n+({1}/{2})$.  

\begin{figure}[h!]
	\centering 
	\includegraphics[width=0.7\linewidth]{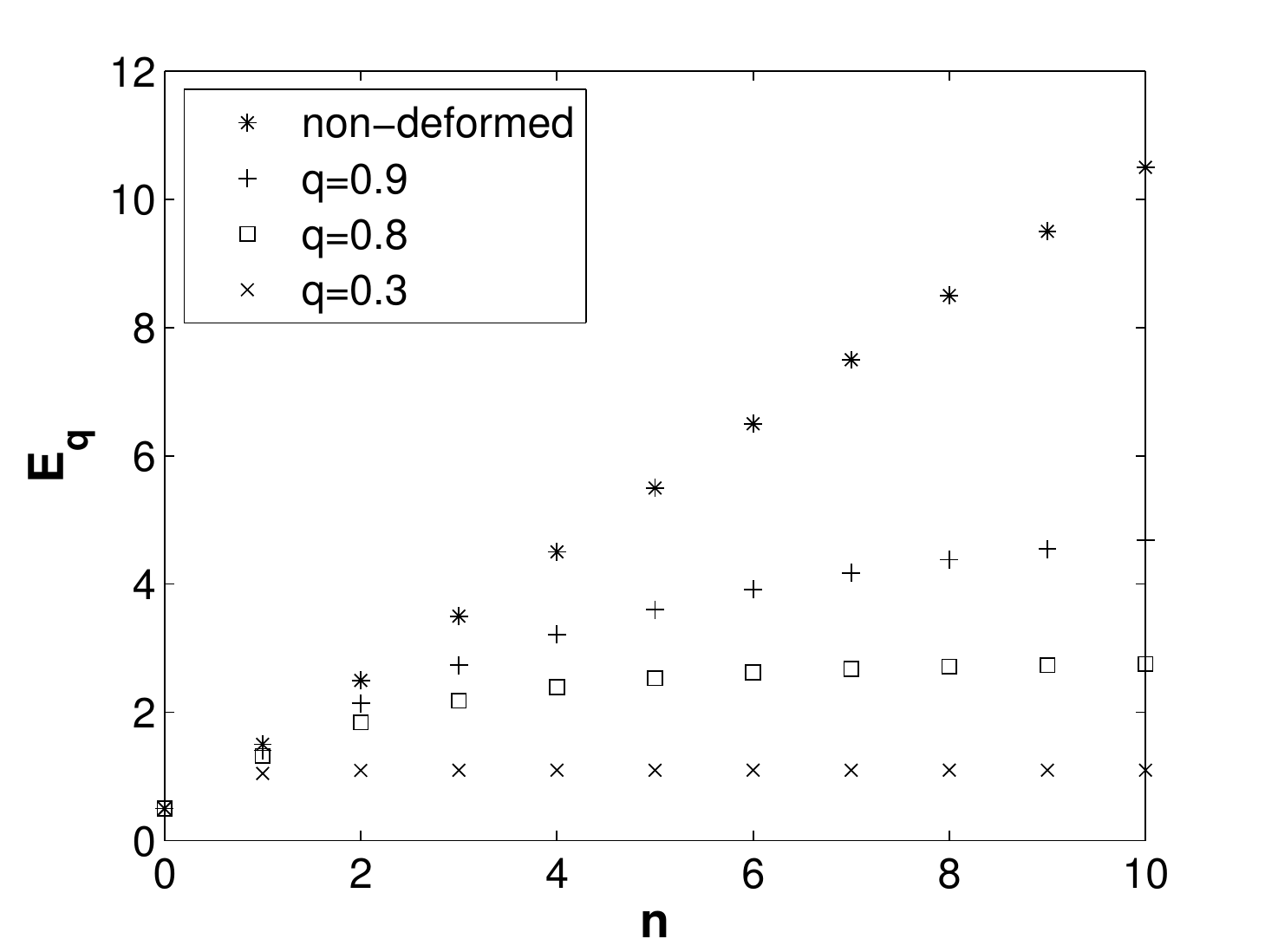}
	\caption{Plot of energy vs $n$ for different $q$ values. The energy eigenvalues are distributed in a non-linear fashion for the $q$-deformed oscillator which is in stark contrast to the linearly distributed energy eigenvalues of the non-deformed harmonic oscillator.}
	\label{energy}  
\end{figure}

When a $q$-deformation is applied to the above system, the Hamiltonian is modified to the form \cite{Eremin}:
\begin{equation} 
H_{q}=\frac{1}{2}( A A^{\dagger}+A^{\dagger}A),
\label{Hamiltonian}
\end{equation} 
where $A$ and its adjoint $A^{\dagger}$ are the deformed annihilation and creation operators, respectively. The oscillator described by the Hamiltonian $H_{q}$ is called the $q$-deformed harmonic oscillator. The operators $A$ and $A^{\dagger}$ obey the deformed commutation relation,
\begin{equation} 
AA^{\dagger}-q^{2}A^{\dagger}A=I, 
\label{Lie}
\end{equation}      
where, $I$ is the identity matrix and $0<q<1$.
The Hamiltonian \eqref{Hamiltonian} has an eigenvalue
\begin{equation}
 E_{q,n}=\left( \left[ n\right] +\frac{q^{2n}}{2}\right),
\end{equation} 
where,
\begin{equation}
\left[n\right]=\frac{1-q^{2n}}{1-q^{2}}.
\label{n} 
\end{equation}
In the limit of $q\rightarrow1$, $\left[n\right]\rightarrow n$ and $E_{q,n}$ reduces to $E_{n}$ of the non-deformed quantum harmonic oscillator. In Fig. \ref{energy}, we can clearly see that the energy values of the $q$-deformed harmonic oscillator depart significantly from the linear energy curve of the non-deformed harmonic oscillator, as mentioned in the previous section. It has been already reported in literature that the non-deformed harmonic oscillator, which has linear dependence of energy eigenvalues in $n$, gives periodic behaviour in the dynamics of expectation values \cite{Sudheesh3}. Thus, the non-linear variation of the energy eigenvalues with $n$ for $q$-deformed harmonic oscillator may give rise to other dynamical properties such as quasi-periodicity and chaos, which is the prime motivation for this study.

The canonical position and momentum operators for the deformed harmonic oscillator have the form \cite{Dey,Jayakrishnan}:
\begin{equation}
X=\dfrac{\sqrt{1+q^{2}}}{2}\left( A^{\dagger}+A\right), 
\label{xt}
\end{equation} 
\begin{equation}
P=i\dfrac{\sqrt{1+q^{2}}}{2}\left( A^{\dagger}-A\right).
\label{pt}
\end{equation}
Further, the action of $A$ and $A^{\dagger}$ on the deformed Fock state $\ket{n}_q$  is defined as \cite{Dey,Jayakrishnan,Eremin},
\begin{equation}
\begin{aligned}
A\ket{n}_{q} &= \sqrt{\left[ n\right] } \ket{n-1}_{q},\\
A^{\dagger}\ket{n}_{q} &=\sqrt{\left[ n+1\right]} \ket{n+1}_{q}.
\end{aligned}
\label{A}
\end{equation}   
 Now, in the next section, we proceed to describe the dynamical evolution of the $q$-deformed observables $X$ and $P$.
\subsection{Time evolved deformed coherent state}
\label{sub2}
\begin{figure}
	\includegraphics[width=0.7\linewidth]{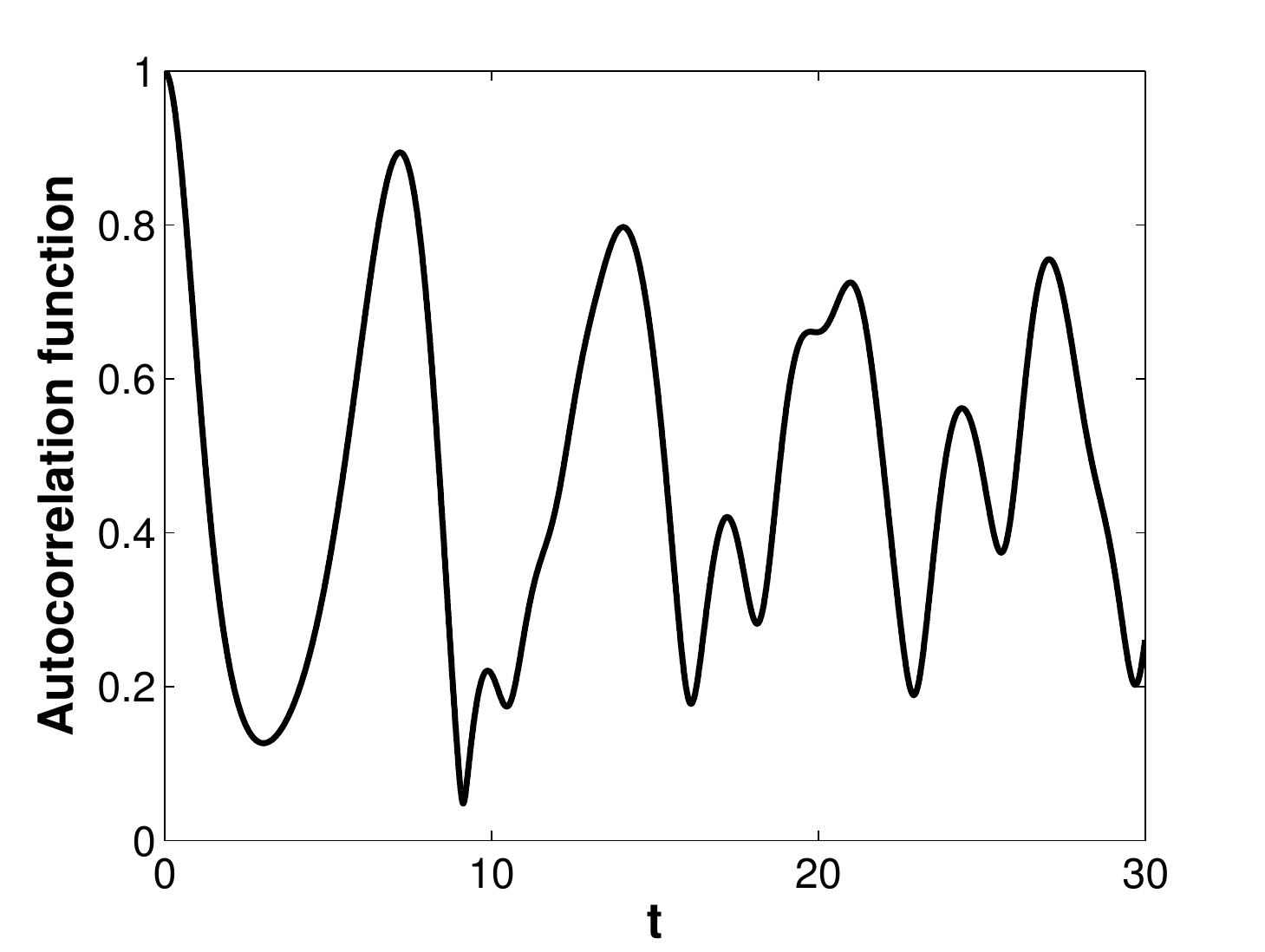}
	\caption{Plot of autocorrelation function as a function of time for $\vert\alpha_{q}\vert^{2}=1$ and $q=0.95$}
	\label{autofig}
\end{figure}

 The $q$-deformed coherent state for the deformed harmonic oscillator is,
\begin{equation}
\ket{\alpha}_{q} = e_{q}^{-\frac{|\alpha_{q}|^{2}}{2}}\sum^{\infty}_{n=0}\frac{(\alpha_{q})^{n}}{\sqrt{\left[ n\right]! }}\ket{n}_{q},
\label{alpha}
\end{equation}
where, $e_{q}^{(\bullet)} = \sum_{n=0}^{\infty}\frac{(\bullet)^{n}}{\left[ n\right]! }$, is the $q$-deformed exponential function.\\ Eq. \eqref{n} can be re-arranged to obtain
\begin{equation}
q^{2n}=1+\left( q^{2}-1\right) \left[ n\right].
\label{q} 
\end{equation}  
Using equations \eqref{alpha} and \eqref{q}, the time evolved  deformed coherent state $\ket{\alpha(t)}_{q}$ is derived as,
\begin{equation}
\begin{aligned}
\ket{\alpha(t)}_{q} &= e^{-iHt}\ket{\alpha(0)}_{q}\; ,\\
&= e_{q}^{-\frac{|\alpha_{q}|^{2}}{2}}\sum_{n=0}^{\infty}\frac{(\alpha_{q}(0))^{n}}{\sqrt{\left[ n\right]!}}e^{-i t\left( \left[ n\right]+ \frac{q^{2n}}{2}\right)  }\ket{n}_{q}\; ,\\
&= e^{-\frac{i t}{2}}e_{q}^{-\frac{|\alpha_{q}|^{2}}{2}}\sum_{n=0}^{\infty}\frac{(\alpha_{q}(0))^{n}}{\sqrt{\left[ n\right]!}}e^{-\frac{i t}{2}\left[ n\right]\left( q^{2}+1\right)  }\ket{n}_{q}\;.
\end{aligned}
\label{alphat}
\end{equation}
From \eqref{alphat}, we obtain the autocorrelation function of the time evolved deformed coherent state: 
\begin{equation}
_{q}\expval{\alpha(0)|\alpha(t)}_{q} = e_{q}^{-|\alpha_{q}|^{2}}e^{\frac{-i t}{2}}\sum_{n=0}^{\infty}\frac{|\alpha_{q}(0)|^{2n}}{\left[ n\right]!}e^{-\frac{i t}{2}\left[ n\right]\left( q^{2}+1\right)}.
\label{auto}
\end{equation}
The corresponding plot is given in Fig. \ref{autofig} from which it is evident that, upon time evolution, the deformed coherent state no longer retains its coherent form.
\par The dynamical evolution of the expectation values of the deformed position operator \eqref{xt} and the deformed momentum operator \eqref{pt} are derived as:
\begin{equation}
\begin{split}
\expval{X(t)}_{q} &= \; _{q}\expval{\alpha(t)| X(t) |\alpha(t)}_{q}\; ,\\
&=\frac{\sqrt{1+q^{2}}}{2} \; e_{q}^{-|\alpha_{q}|^{2}}\\
& \times \; \sum_{n=0}^{\infty}\left\lbrace \frac{|\alpha_{q}|^{2n}}{\left[ n\right]!}\; \alpha_{q} \; e^{\frac{i t}{2}\left( 1+q^{2}\right) \left( \left[ n\right] - \left[ n+1\right] \right) }\right. \\ 
&+ \left. \frac{|\alpha_{q}|^{2n+1}}{\left[ n\right]!} \; \alpha_{q}^{-1} \; e^{-\frac{it}{2}\left( 1+q^{2}\right)\left( \left[ n\right] - \left[ n+1\right]  \right)  }\right\rbrace. 
\end{split}
\end{equation}
\begin{equation}
\begin{split}
\expval{P(t)}_{q} &= \; _{q}\expval{\alpha(t)| P(t) |\alpha(t)}_{q}\; ,\\
&=i \; \frac{\sqrt{1+q^{2}}}{2} \; e_{q}^{-|\alpha_{q}|^{2}}\\ 
&\times \; \sum_{n=0}^{\infty}\left\lbrace \frac{|\alpha_{q}|^{2n+1}}{\left[ n\right] !} \; \alpha_{q}^{-1} \; e^{-\frac{i t}{2}\left( 1+q^{2}\right)\left( \left[ n\right] - \left[ n+1\right]  \right)} \right. \\ 
&- \left. \frac{|\alpha_{q}|^{2n}}{\left[ n\right]!} \; \alpha_{q} \; e^{\frac{i t}{2}\left( 1+q^{2}\right) \left( \left[ n\right] - \left[ n+1\right] \right) }\right\rbrace\; ,
\end{split}
\end{equation}  
using the expressions in equations \eqref{A} and \eqref{alphat}.
\subsection{Nature of the time evolved observables}
\label{sub3}
The time evolved observables, in this case, the expectation values of the deformed position and momentum, produce respective time series when computed numerically. One can perform a generic analysis of this series by applying the common procedures followed for a typical time series data. Here, the nature of these time evolved variables are studied using four complimentary analysis tools, namely, first-return-time distributions, recurrence plots, Lyapunov exponents and the power spectra. 
\subsubsection{Recurrence plots:}
Recurrence plot is defined as the graphical representation of 
\begin{equation}
R_{i,j}=
\begin{cases}
1, &\text{if $\vec{x_{i}}\approx\vec{x_{j}}$;}\\
0, &\text{otherwise.}
\end{cases}
\quad i,j=1,2,...N,
\end{equation}
where $N$ is the number of data points under consideration and $\vec{x_{i}}\approx\vec{x_{j}}$ refers to its equivalence within a designated parameter $\epsilon$ \cite{Marwan}. The simulation of this calculation produces an $N\times N$ matrix, whose elements are a series of ones and zeroes, with $1$ representing points that lie close to each other. We can classify a given series as periodic, quasi-periodic or chaotic based on the particular features of its recurrence plots. As a general rule, we may establish that periodic trajectories are characterized by parallel, equidistant diagonal lines. Quasi-periodicity is characterised by two or more sets of parallel, diagonal lines. Chaos is characterized by a single \emph{line of identity} which may or may not be surrounded by other short broken lines at random distances from the \emph{line of identity}.

\subsubsection{Power spectra}
We also utilize the power spectrum of the time series to understand the nature of the non-linear system better. The power spectra are easily obtained by the technique of fast Fourier transform of the time series. It is to be noted that in case of a chaotic series, the power spectrum displays ``grassiness" with the spectrum also showing a decreasing trend. Quasi-periodicity is indicated by peaks which may or may not exhibit splitting \cite{Sudheesh,Dumont}.

\subsubsection{First-return-time distributions:} 
 First-return-time distribution encompasses information about the recurrence of a small range of values over a large series of datapoints \cite{Sudheesh,Sudheesh2}. We construct computational cells of suitable sizes and determine the frequency of recurrence of datapoints within this cell. We compute the probability of recurrence and attempt to fit appropriate probability distributions. Based upon the requirement of ergodicity, from the Poincar\'{e} recurrence theorem \cite{Baxter}, the recurrence time $t$ is found to satisfy an exponential distribution of the form $F_{1}(\tau)=({1}/{\tau}) \; e^{{-t}/{\tau}}$, where the mean $\mu$ is given by $\mu^{-1}=\expval{\tau}$ \cite{Sudheesh,Baxter}. This implies that the first-return-time distributions which satisfy this relation correspond to ergodic behaviour. In the present study, we obtain the first-return-time distribution for cell sizes less than or equal to $10^{-3}$.\\
 
\subsubsection{Lyapunov exponent}
The next parameter analysed in this study is the Lyapunov exponent. The Lyapunov exponent ($\lambda$) describes the divergence from an initial trajectory, of an almost identical trajectory produced by an infinitesimal perturbation in the initial conditions. A positive Lyapunov exponent is indicative of chaos in the system, in which case the trajectories diverge at an exponential rate defined by the largest Lyapunov exponent $\lambda_{max}$ of the system \cite{Rosenstein}. In particular, the slope of the graph of $ln(d(j))$ vs $j\bigtriangleup t$, where, $d(j)$ stands for the divergence of the trajectory and $j\bigtriangleup t$ denotes the time, gives us an idea of the sign of the Lyapunov exponent and its magnitude. In this study, we use the Rosenstein algorithm \cite{Rosenstein} to determine the greatest Lyapunov exponent $\lambda_{max}$. We also cross verify our calculation of the Lyapunov exponent using another conventional algorithm called the Wolf algorithm \cite{Wolf}. The results were found to be in agreement with each other.\\

It must be noted that all these analysis methods ideally complement each others findings and none of them can be utilized as a confirmation for a system property, on its own.
\section{Results and Discussion}
\label{sec3}
\subsection{Analysis of dynamical variations} 
\label{sub31}
 As a preliminary step, we analyse the autocorrelation function in Fig. \ref{autofig}. Here we present the results obtained for real values of $\alpha_{q}$ but our result holds true for any general complex value of $\alpha_{q}$. If a system is periodic, the respective autocorrelation function is bound to be periodic. For $\alpha_{q}=1$ and $q=0.9$, the autocorrelation function appears to peak at random intervals, hinting at the probable chaotic nature of the system for the chosen set of parameters. Based on this observation, we proceed to investigate further into the properties of the system.
 
 A qualitative reasoning for the system at hand can be obtained from the plots of its dynamic evolution and phase space diagrams.
\begin{figure}[h!]
	\centering 
	\subfloat[$q$=0.1]{\includegraphics[width=0.5\linewidth]{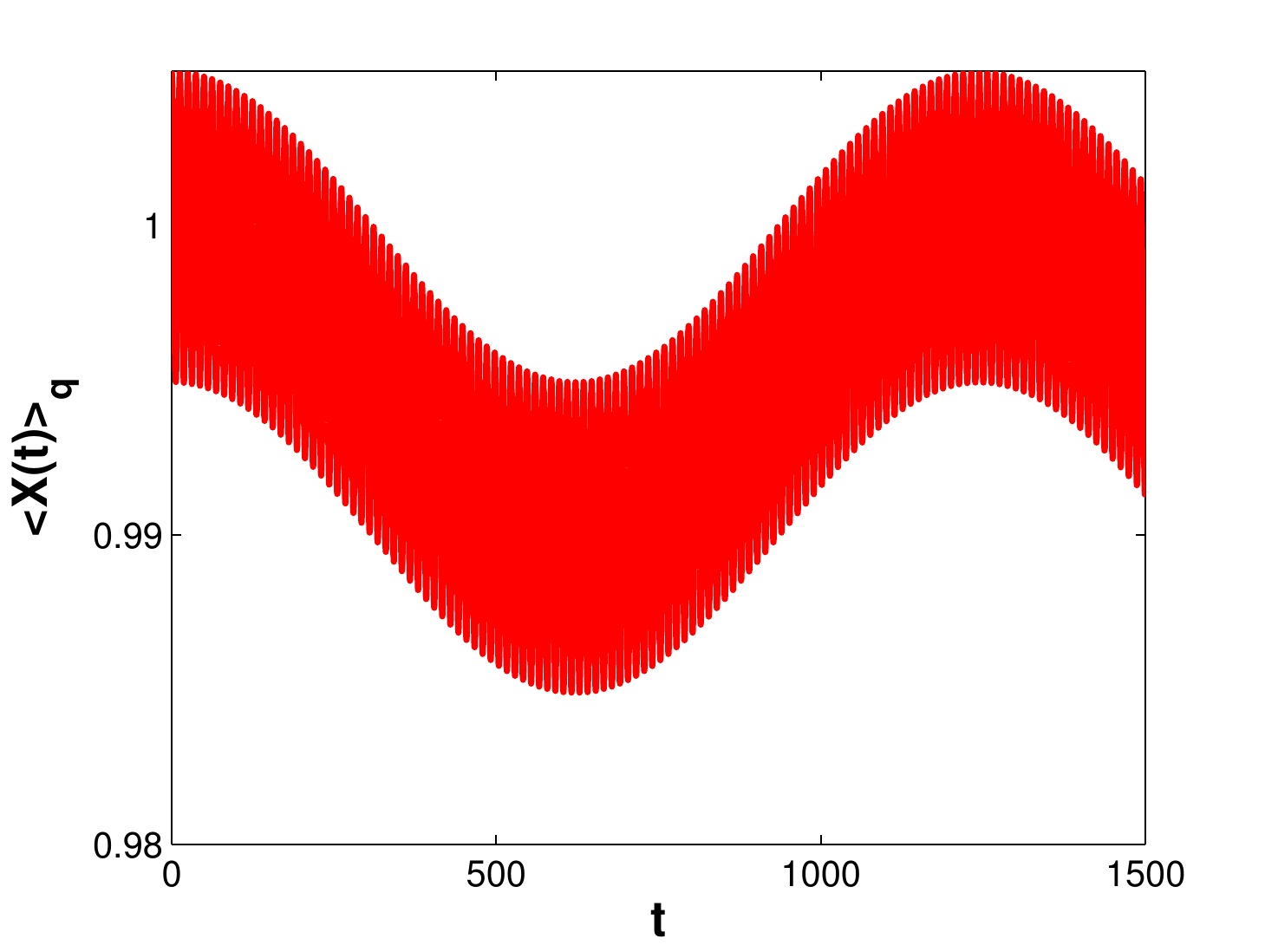}}
	\subfloat[$q$=0.1]{\includegraphics[width=0.5\linewidth]{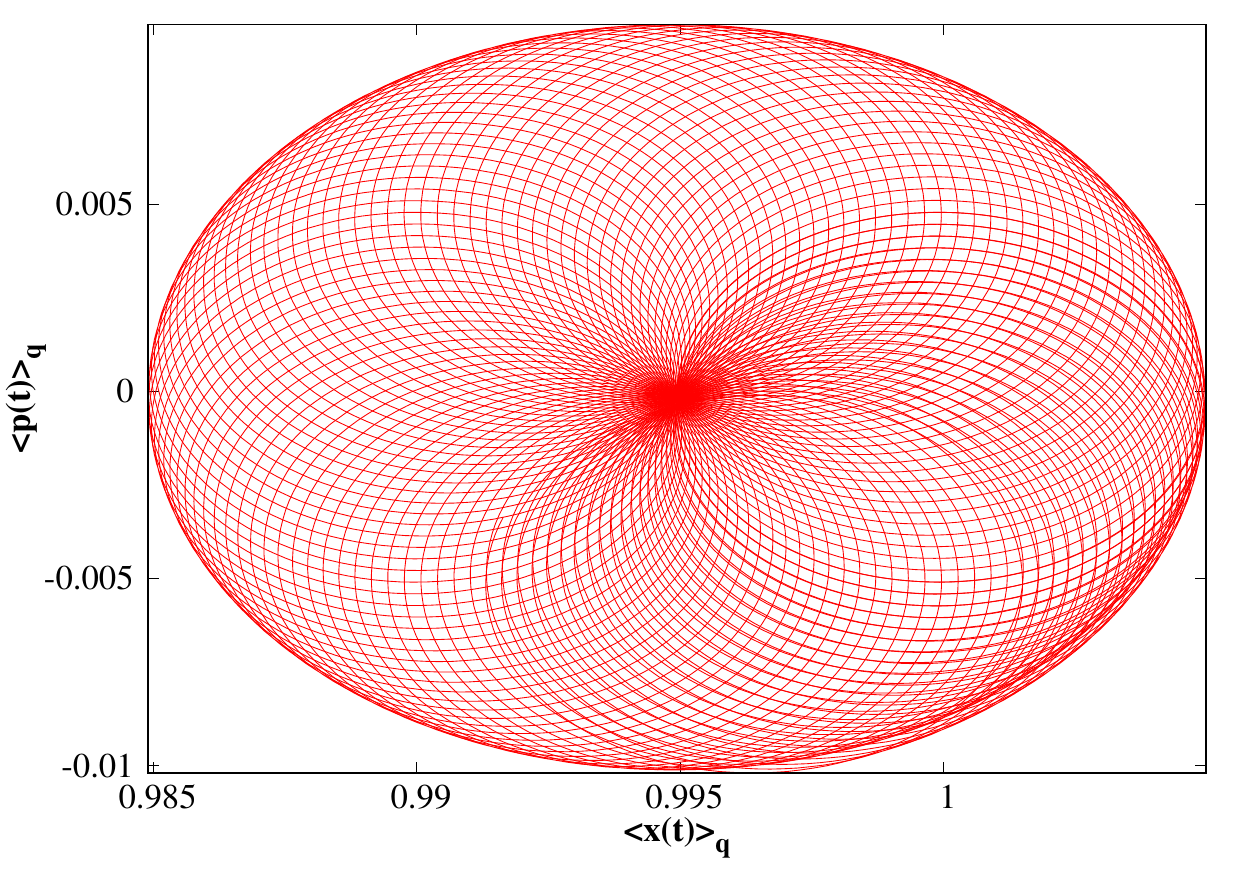}}\\
	\subfloat[$q$=0.2]{\includegraphics[width=0.5\linewidth]{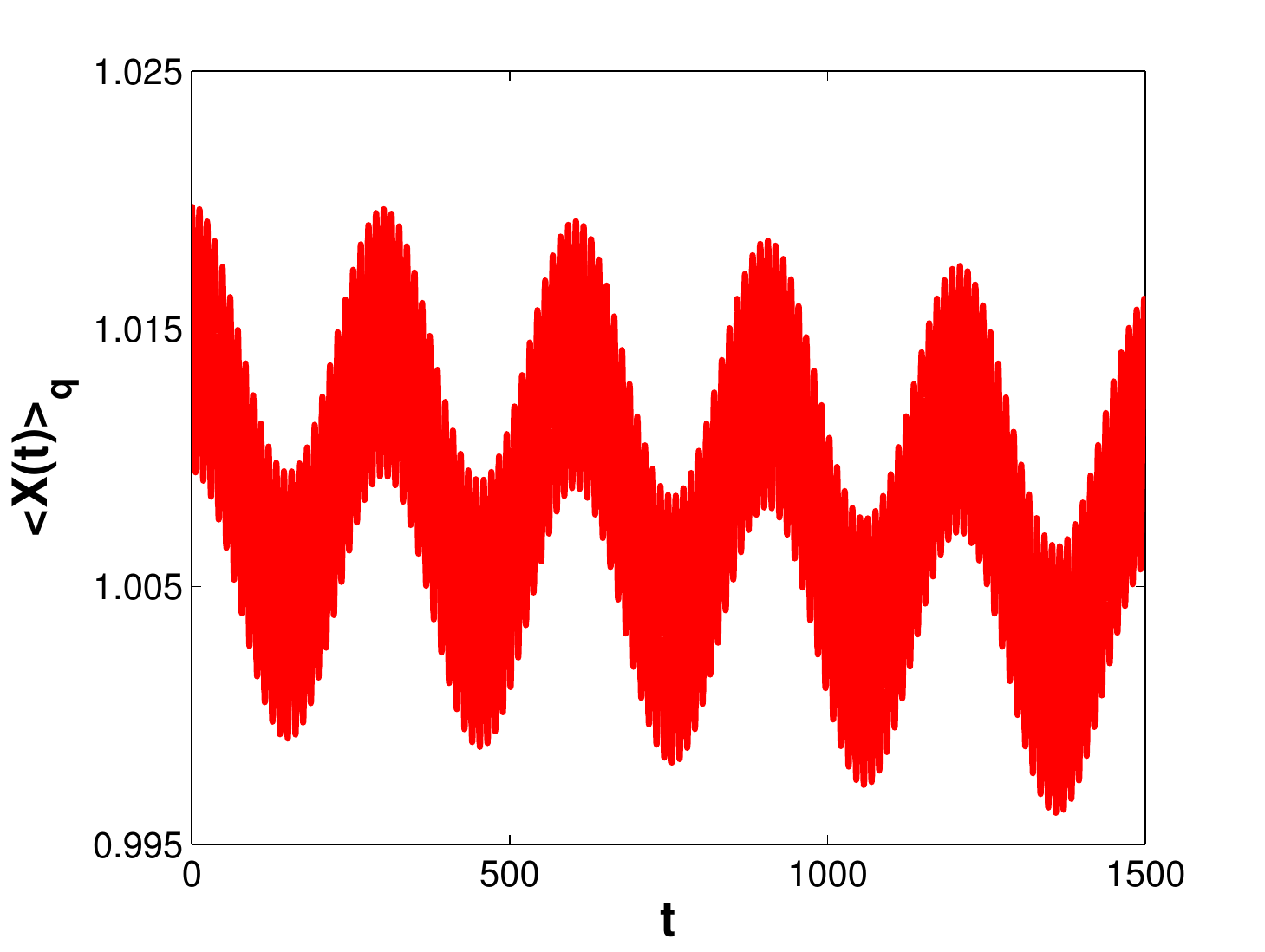}}
	\subfloat[$q$=0.2]{\includegraphics[width=0.5\linewidth]{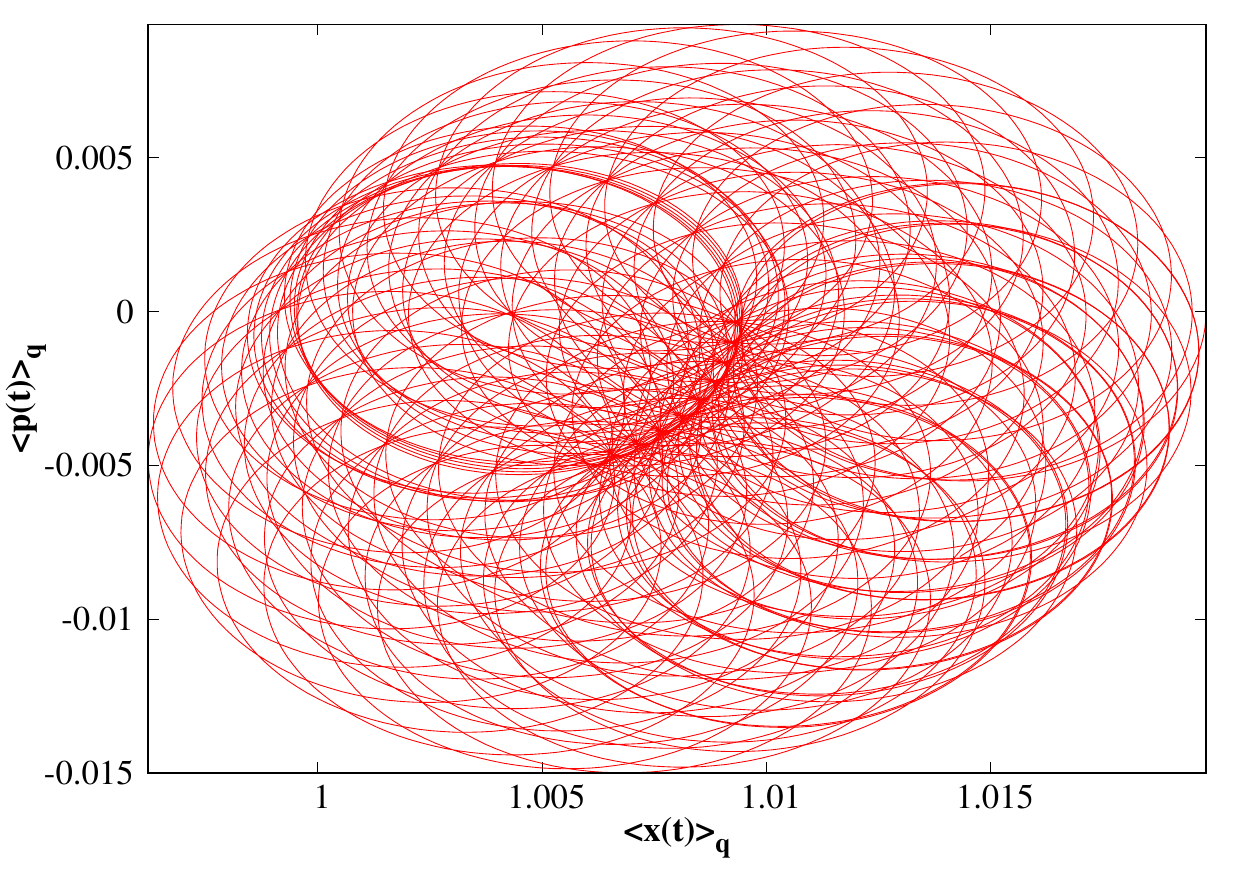}}\\
	\subfloat[$q$=0.9]{\includegraphics[width=0.5\linewidth]{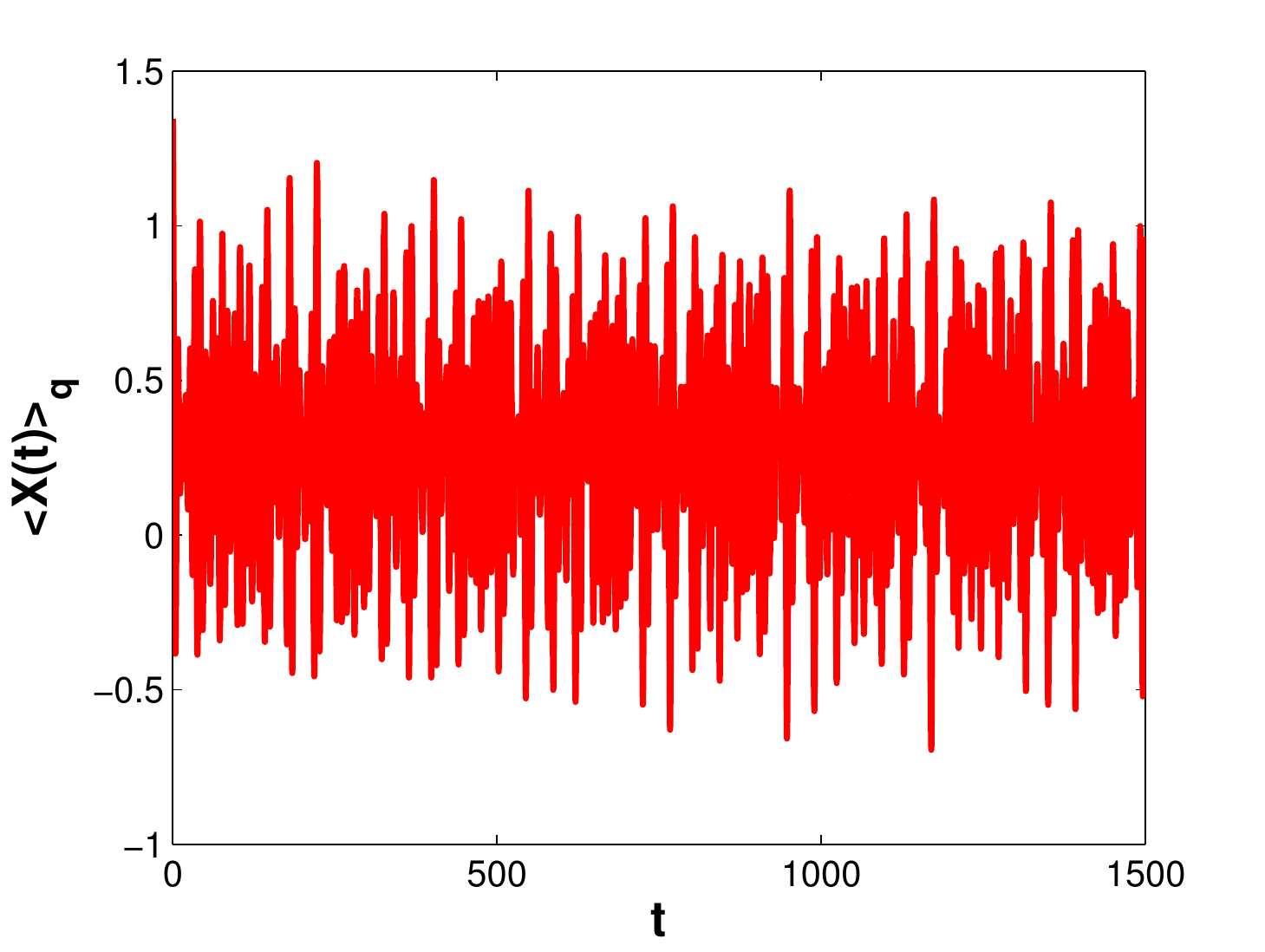}}
	\subfloat[$q$=0.9]{\includegraphics[width=0.5\linewidth]{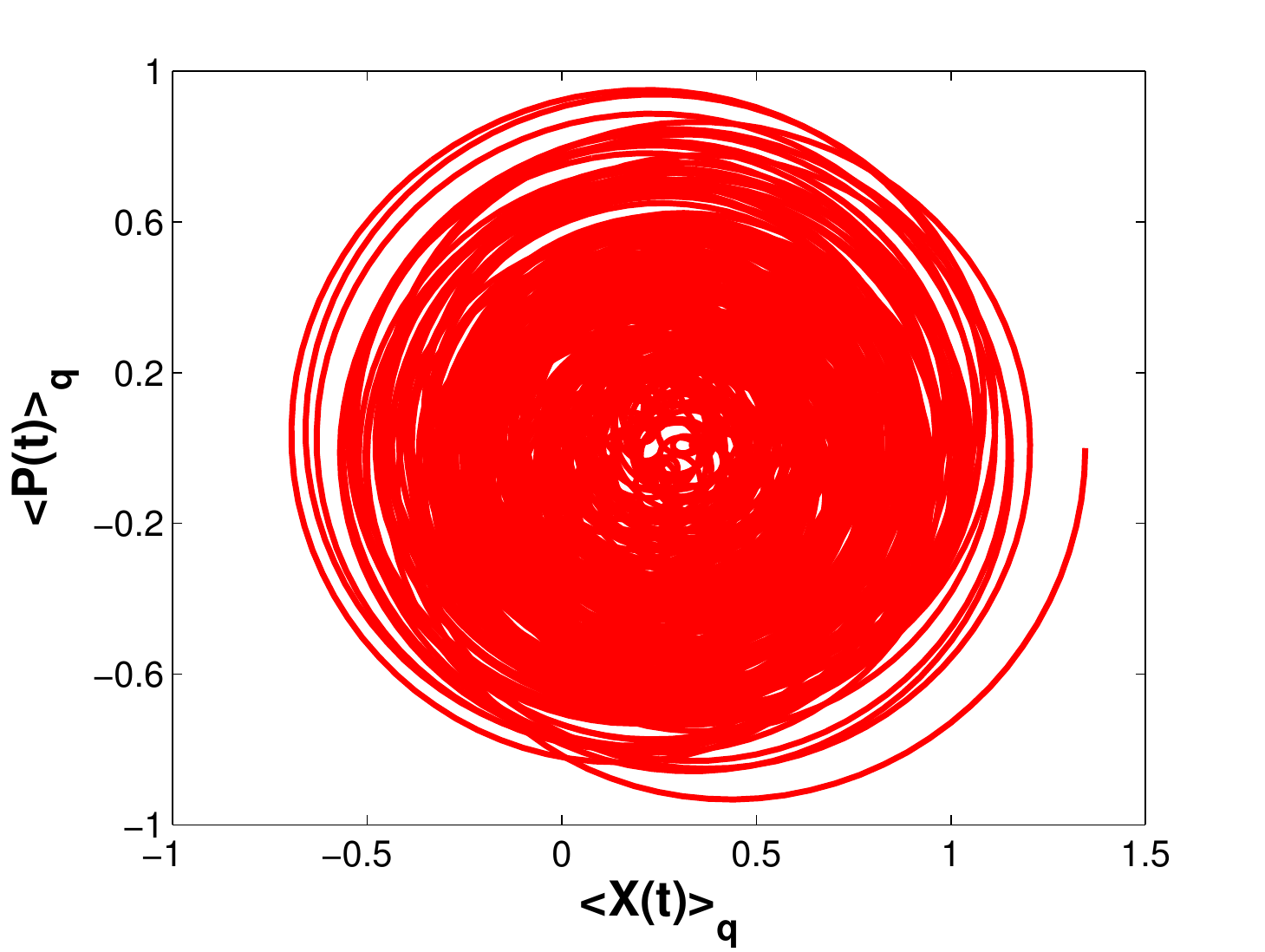}}\\
	\subfloat[$q$=0.99]{\includegraphics[width=0.5\linewidth]{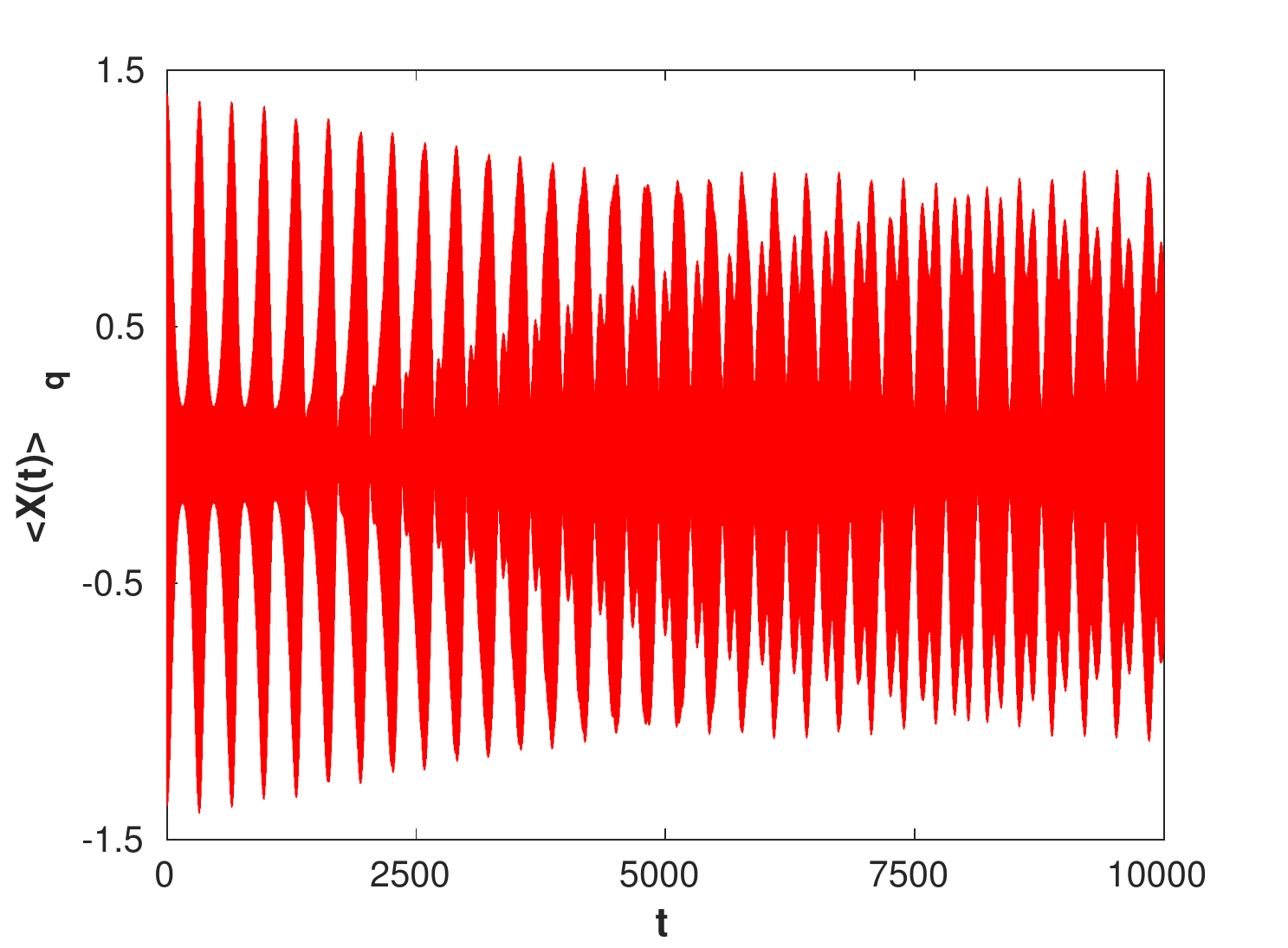}}
	\subfloat[$q$=0.99]{\includegraphics[width=0.5\linewidth]{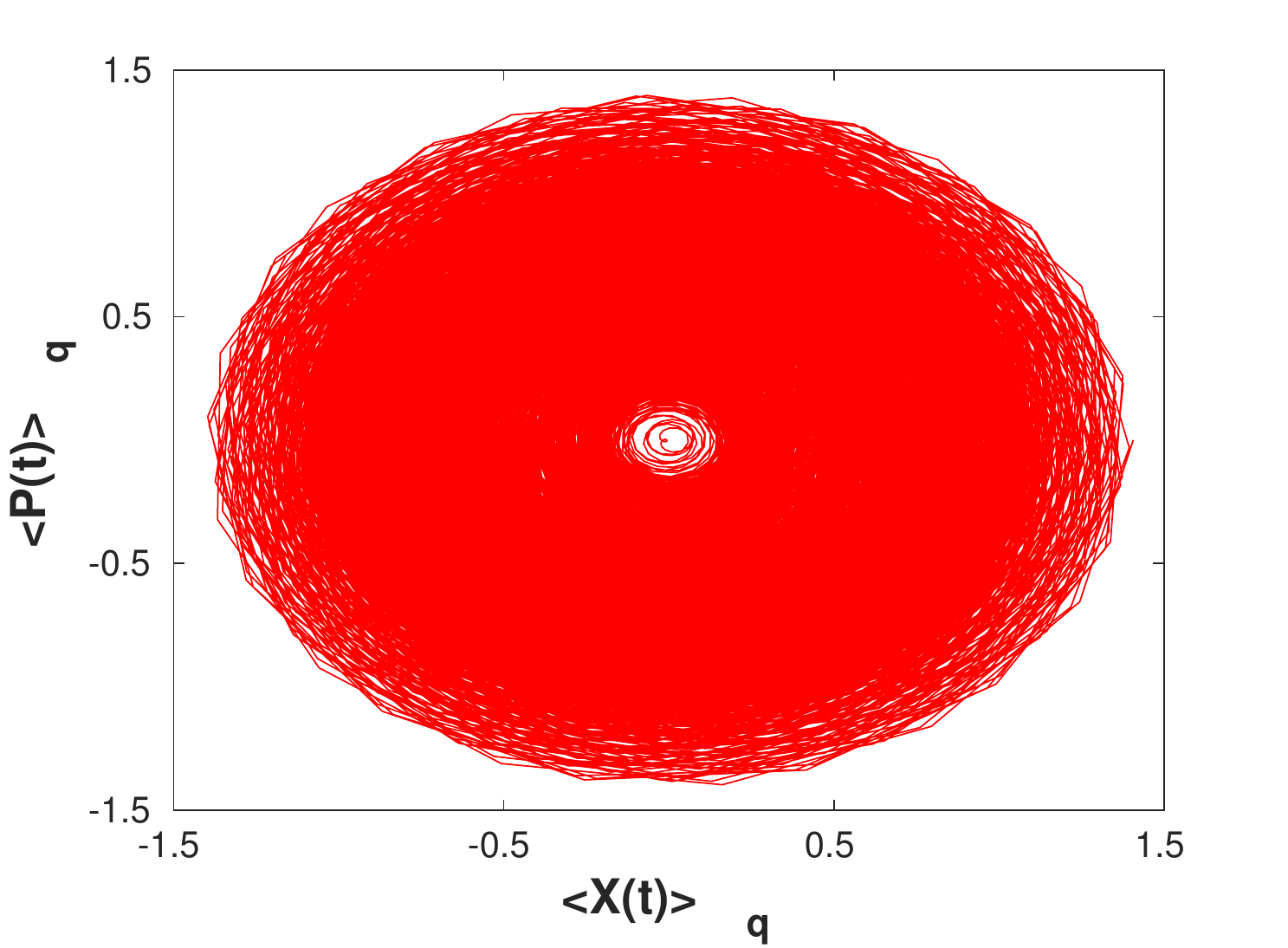}}
	\caption{Time evolution of $\expval{X(t)}_{q}$ (on the left column) and corresponding phase space diagrams $\expval{X(t)}_{q}$ vs $\expval{P(t)}_{q}$ (on the right column), for different $q$ values and $\alpha_{q}=1$.} 
	\label{xtvst}
\end{figure}
These results, shown in Fig. \ref{xtvst}, reinforce our guesses regarding the chaotic nature of the expectation values. Similar plots have been obtained previously for physics-type $q$-deformation in \cite{Buzek}, but an analysis in terms of chaos was not performed on the data in their study. Now, in the present case, an interesting trend is observed when the values of $\alpha_{q}$ and $q$ are varied. In Fig. \ref{xtvst}, we summarize how the dynamic behaviour of the expectation values vary gradually with $q$ for $\alpha_{q}=1$. In Fig. \ref{xtvst}(g) and Fig. \ref{xtvst}(h) we plot the data for a longer time $(10^{4})$ to make its nature more clear. All other plots in Fig. \ref{xtvst} have been made for $t=1500$ as the nature of the system in these cases is clearly evident with this data set. On the higher extreme (as $q$ increases/ as deformation decreases), we see that the behaviour gradually approaches that of a harmonic oscillator, while on the lower end (as $q$ decreases/ as deformation increases), the behaviour first appears chaotic, turns quasi-periodic and finally progresses towards periodicity. Specifically, this new found periodicity seems to emerge for values of $q\leq0.1$. As $q$ increases for the same $\alpha_{q}$, the plot of expectation values vs $t$ appears to get constricted between the given pair of points. This constriction increases relatively with every step rise in $q$. This gradual change appears to be taking the system from periodic to aperiodic behaviour. The corresponding behaviour is reflected in the phase space diagram by the effective ``order" or ``disorder" of the data points. However, the system does not continue to be aperiodic, rather as $q\rightarrow 1$, its behaviour rapidly approaches that of the non-deformed harmonic oscillator. We study these characteristics of the expectation values in detail in the following sections quantitatively. 
\begin{figure}[h!]
	\centering
	\subfloat[$\alpha_{q}$=0.1]{\includegraphics[width=0.5\linewidth]{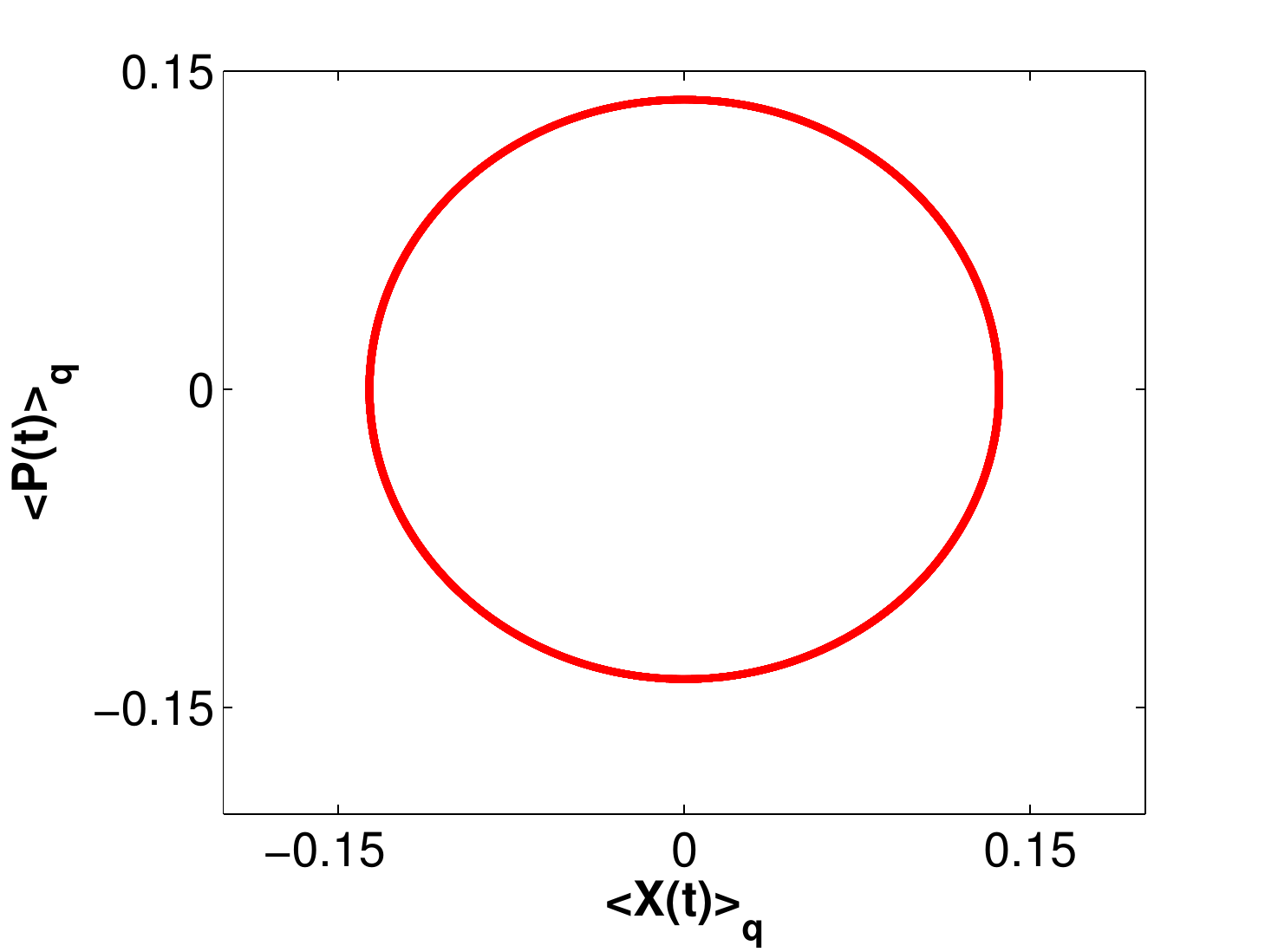}}
	\subfloat[$\alpha_{q}$=0.3]{\includegraphics[width=0.5\linewidth]{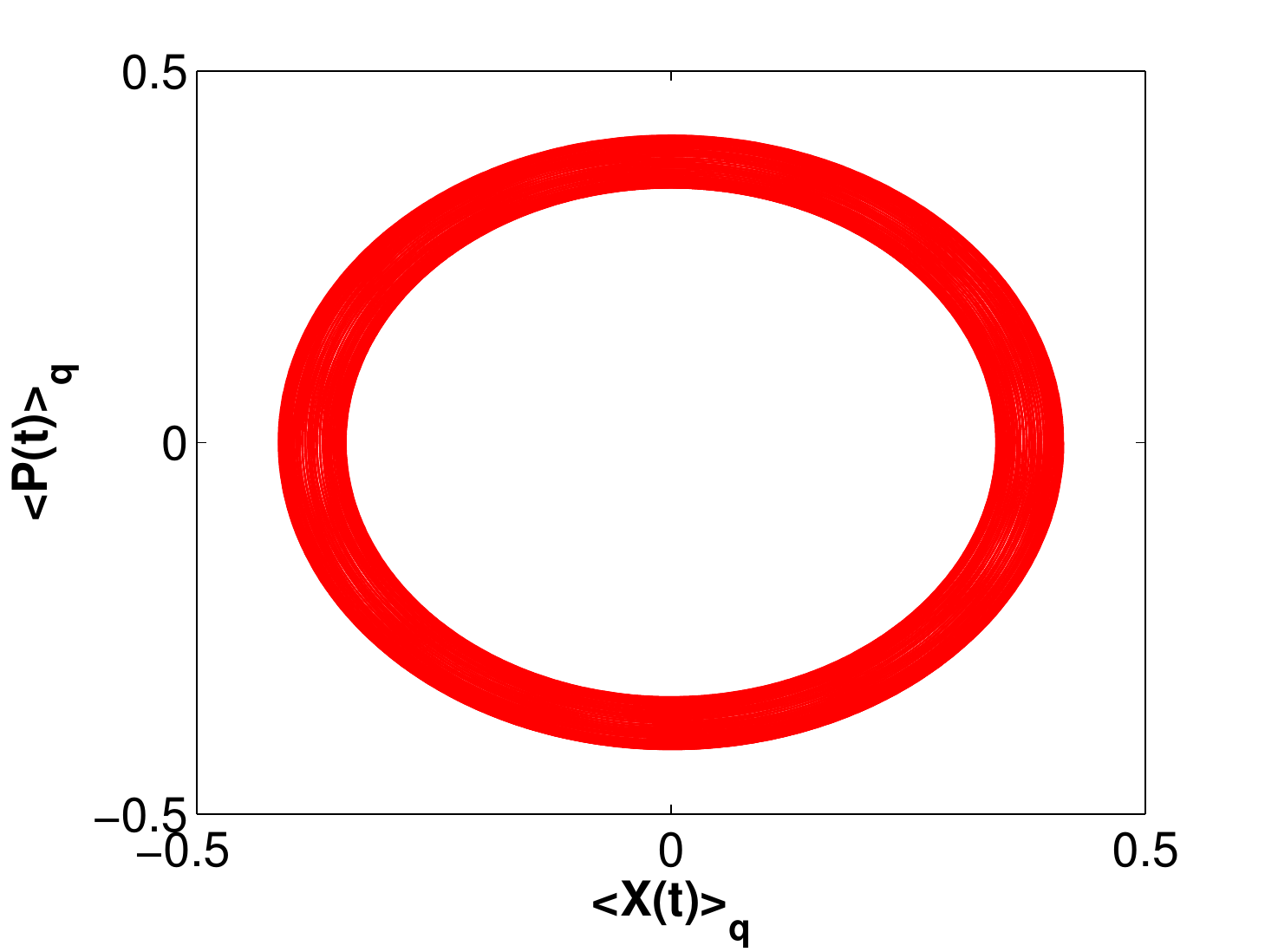}}\\
	\subfloat[$\alpha_{q}$=0.5]{\includegraphics[width=0.5\linewidth]{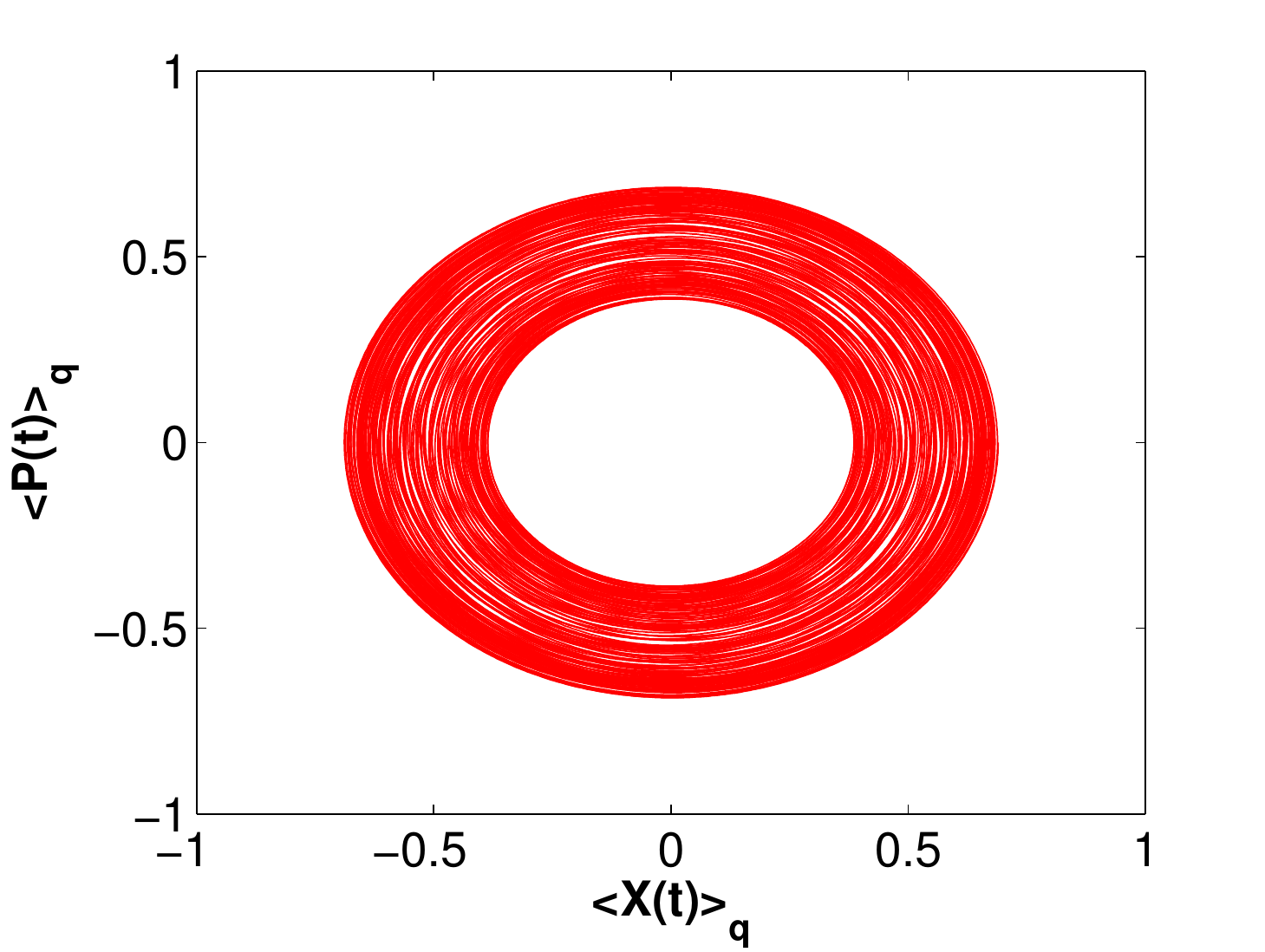}}
	\subfloat[$\alpha_{q}$=0.7]{\includegraphics[width=0.5\linewidth]{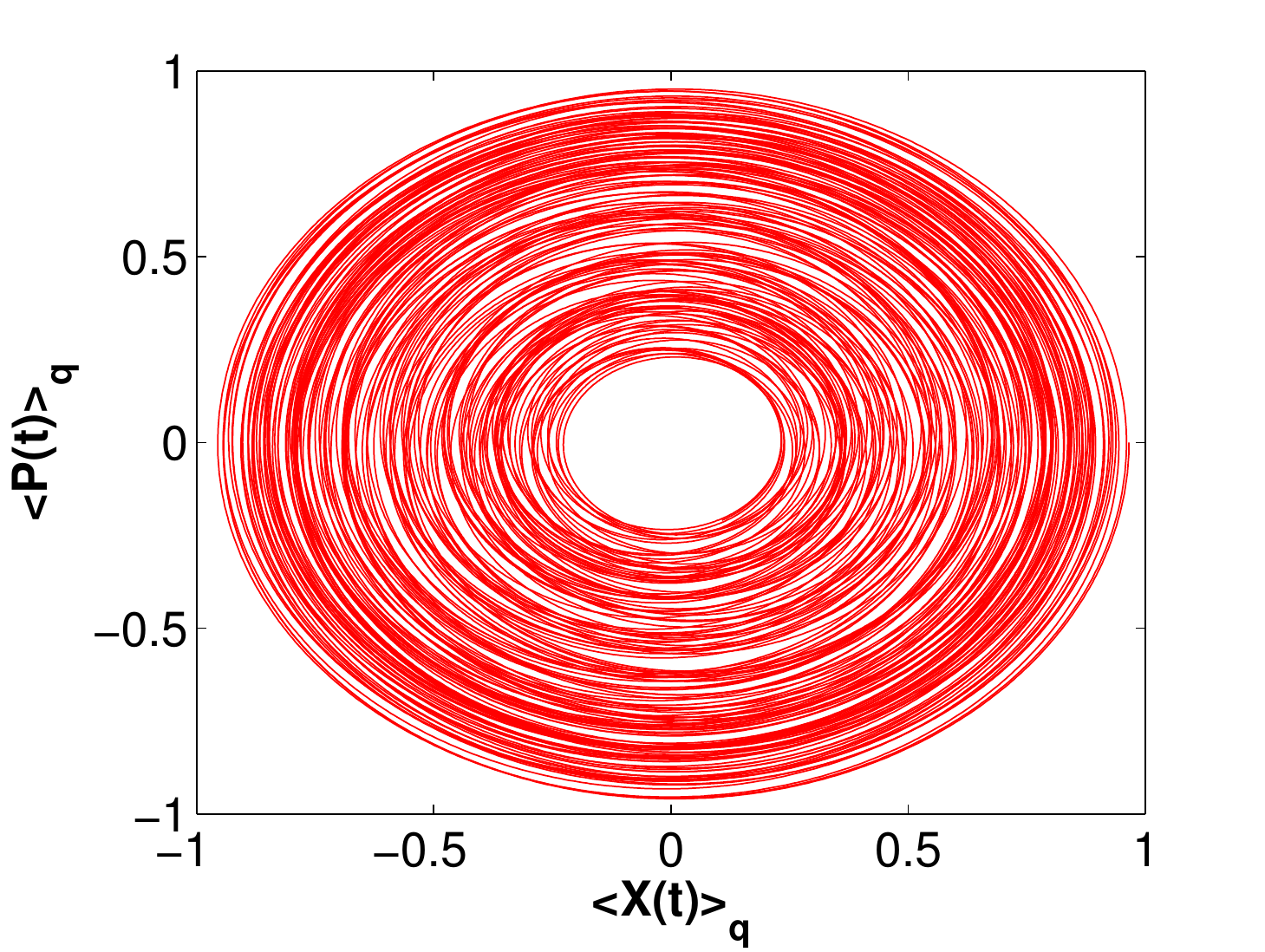}}
	\caption{Phase space diagrams for different values of $\alpha_{q}$ and q=0.95.} 
	\label{xvsp} 
\end{figure}
\par Another interesting dependence of the system dynamics is on the parameter $\alpha_{q}$. In Fig. \ref{xvsp}, the phase space plots show specific behaviour depending on $\alpha_{q}$ value. We identify that the typical chaotic behaviour exists only for $\alpha_{q} >0.5$ when $q=0.95$. Below this value of $\alpha_{q}$, the phase space plot takes on a band-like structure. In the later sections, we will explain how these band-structures conform to quasi-periodic behaviour. The width of the band is seen to decrease with $\alpha_{q}$. For lower values of $\alpha_{q}$ (typically for $|\alpha_{q}|^{2}\leq 0.1$), the behaviour approaches periodicity. It should be noted that there exists a restriction on $\alpha_{q}$ values for a given value of $q$ \cite{Eremin}:
\begin{equation} 
 |\alpha_{q}|^{2}\leq\frac{1}{1-q}.
 \label{limit}
\end{equation}
When $\alpha_q=2$, due to the limit in equation \eqref{limit}, the values of $q$ below 0.75 are not allowed. For $\alpha_{q}=1$, \eqref{limit} permits all $q$ values in the range $0\leq q\leq1$. Now let us look at some more features of the system which will help us to get a better qualitative understanding of the properties stated above.
\subsection{Recurrence plots}
\par
\begin{figure}
	\centering
	\subfloat[$q$=0.1]{\includegraphics[width=0.55\linewidth]{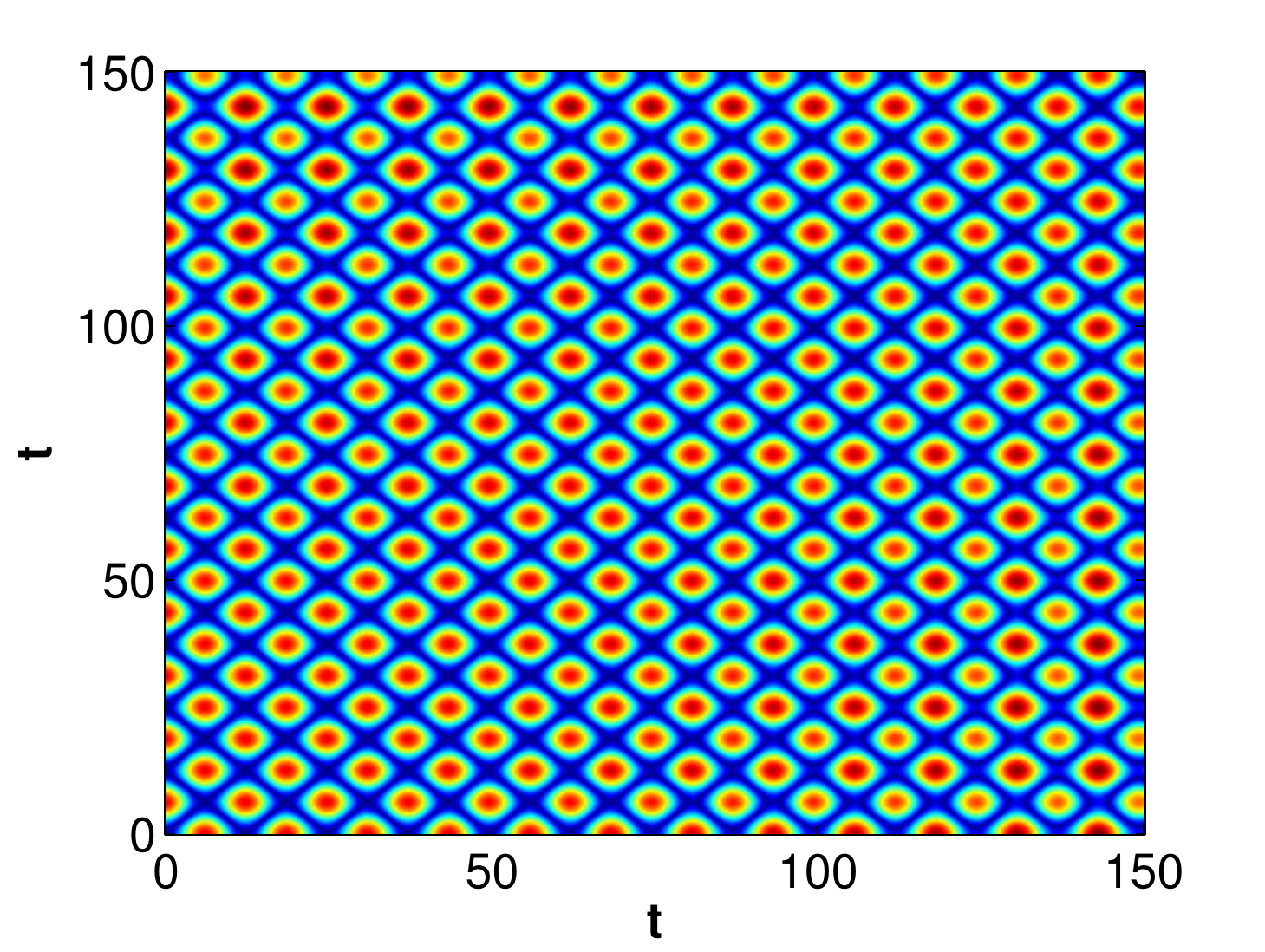}}
	\subfloat[$q$=0.2]{\includegraphics[width=0.55\linewidth]{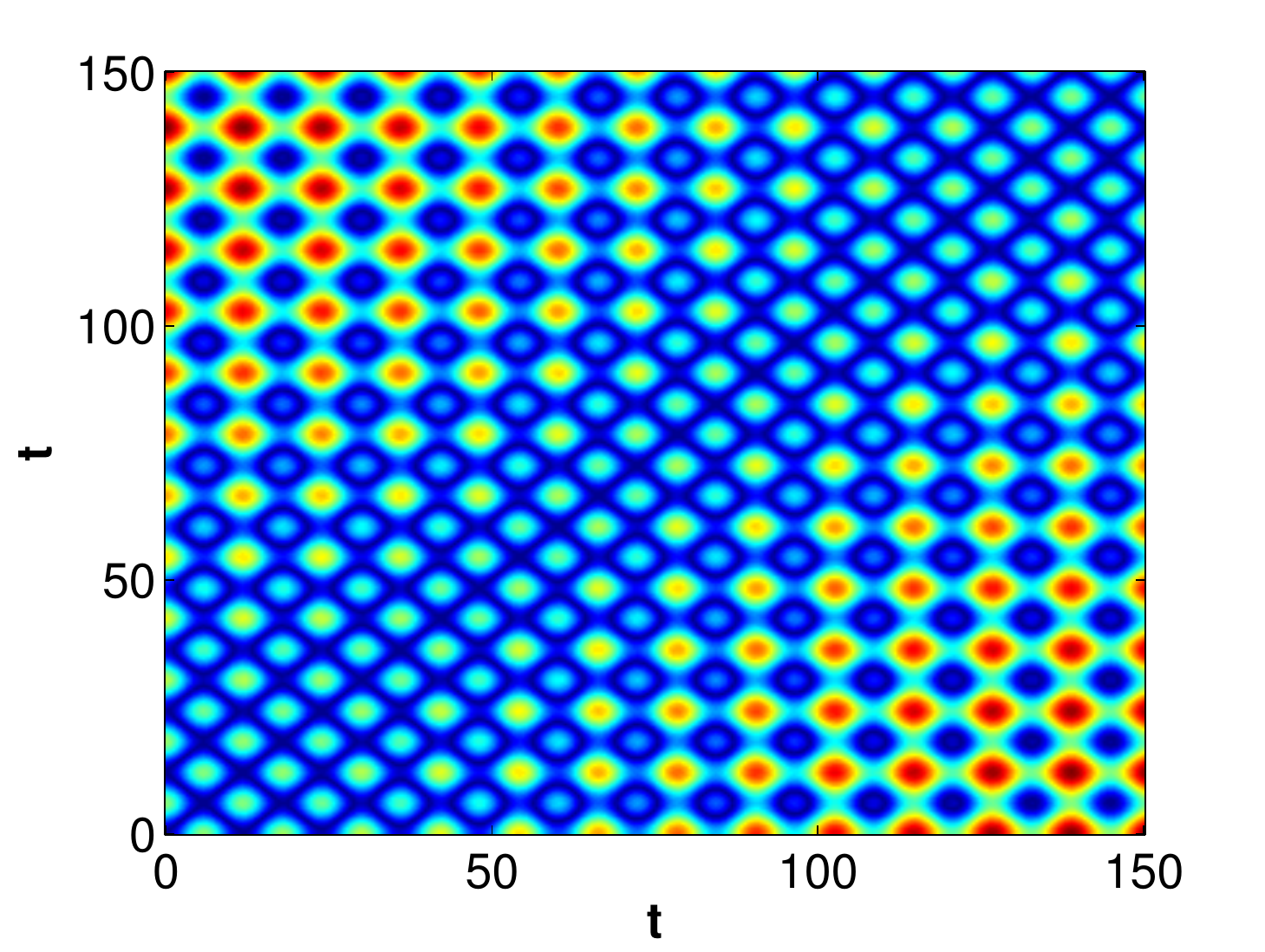}}\\
	\subfloat[$q$=0.9]{\includegraphics[width=0.55\linewidth]{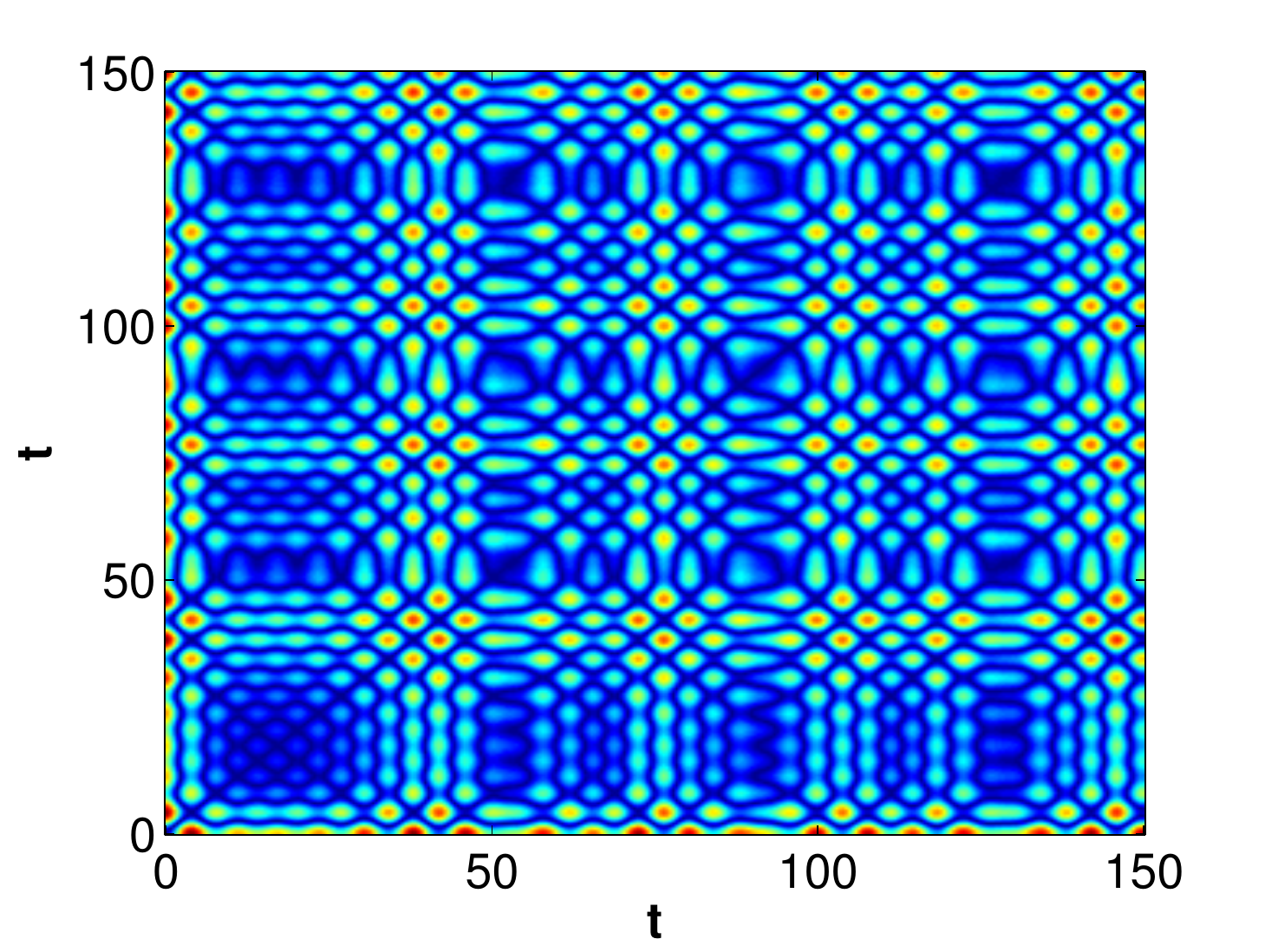}}	
	\subfloat[$q$=0.99]{\includegraphics[width=0.55\linewidth]{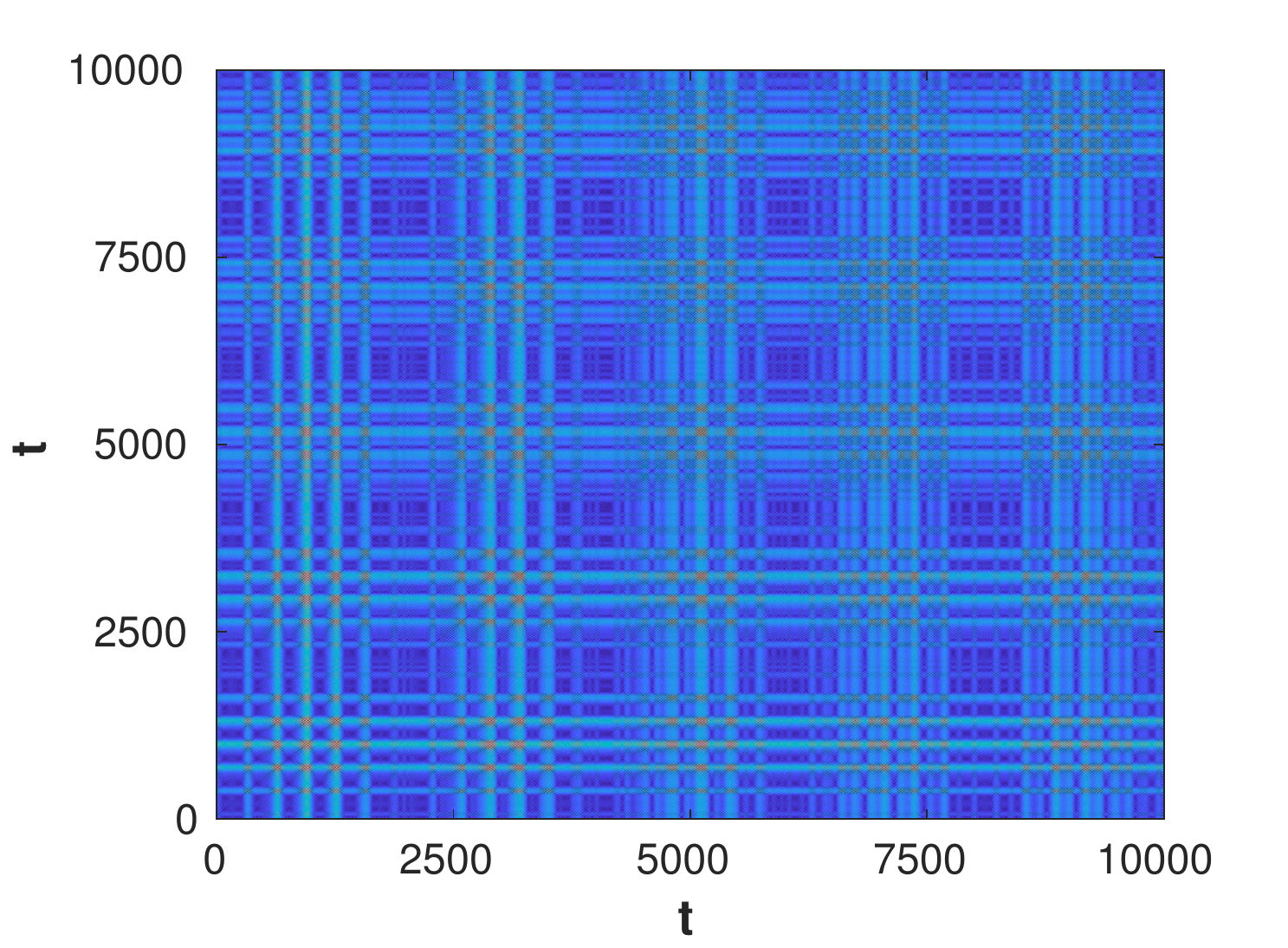}}
	\caption{Recurrence plots for different $q$ values when $\alpha_{q}=1$. (a) corresponds to periodic data, (b), (d) correspond to quasi-periodic data and (c) corresponds to chaotic data.}
	\label{Reca1}
\end{figure}

 The recurrence plots in Fig. \ref{Reca1}, underscores our observations regarding the time series. Through careful observation, we identify the defining features in the recurrence plots that classify its behaviour as periodic or aperiodic.

In the Fig. \ref{Reca1}, when $\alpha_q=1$ for $q=0.1$, the recurrence plot is characterised by equidistant, parallel diagonals, which is an attribute of periodic systems. For $q=0.2$, we can roughly identify two sets of parallel lines, each set composed of respective equidistant lines. This denotes the presence of two periods for the corresponding time series, thus implying a quasi-periodic behaviour. In Fig. \ref{Reca1}(c), we can clearly distinguish a \textit{line of identity} (as described in the previous section), combined with symmetrically distributed short broken lines at random distances on both of its sides. At $q=0.99$, the system displays a quasi-periodic behaviour and progresses towards periodicity as $q\rightarrow1$. Thus, we are able to mark out a clear chaotic regime with the help of these recurrence plots.   

\subsection{Power Spectrum}
\begin{figure}[h!]
	\centering
	\subfloat[$q$=0.9]{\includegraphics[width=0.5\linewidth]{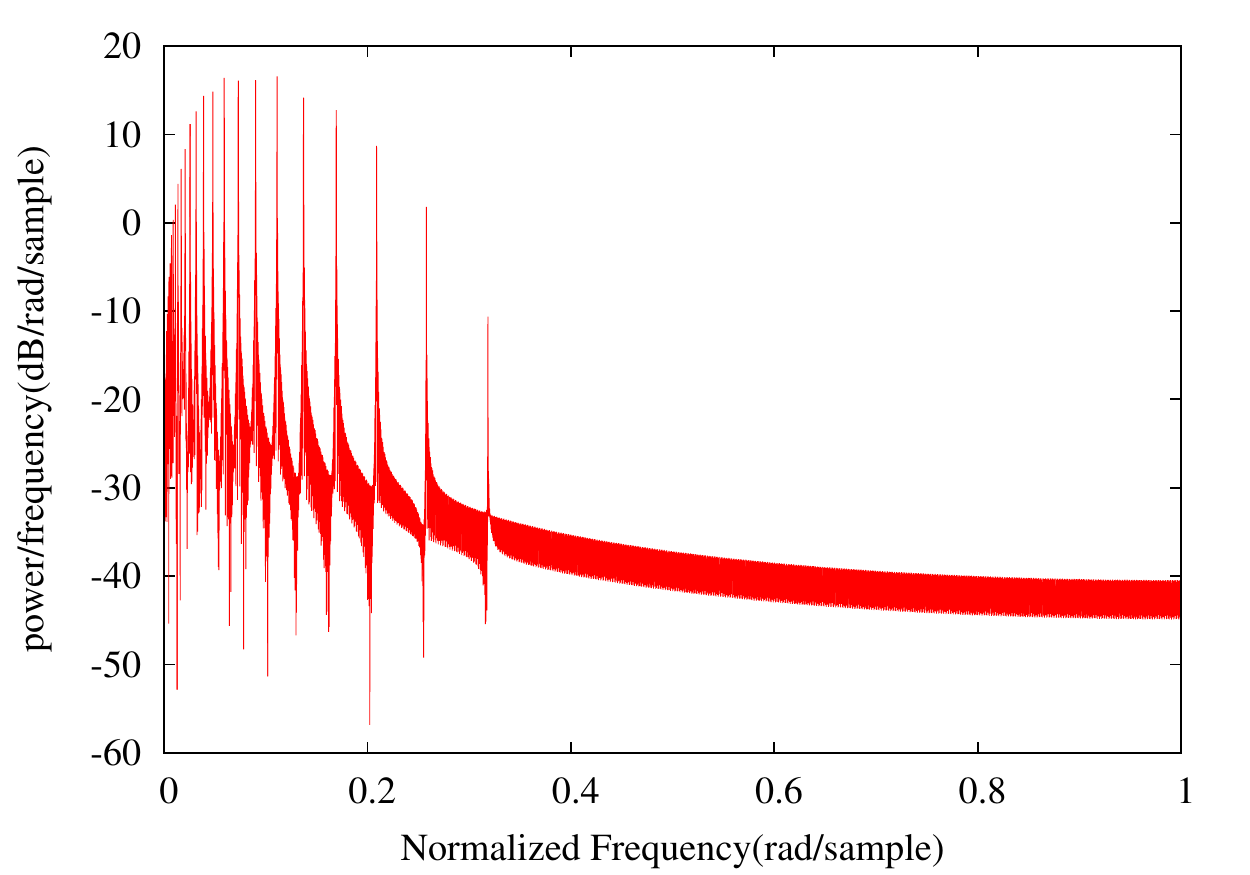}}
	\subfloat[$q$=0.999]{\includegraphics[width=0.5\linewidth]{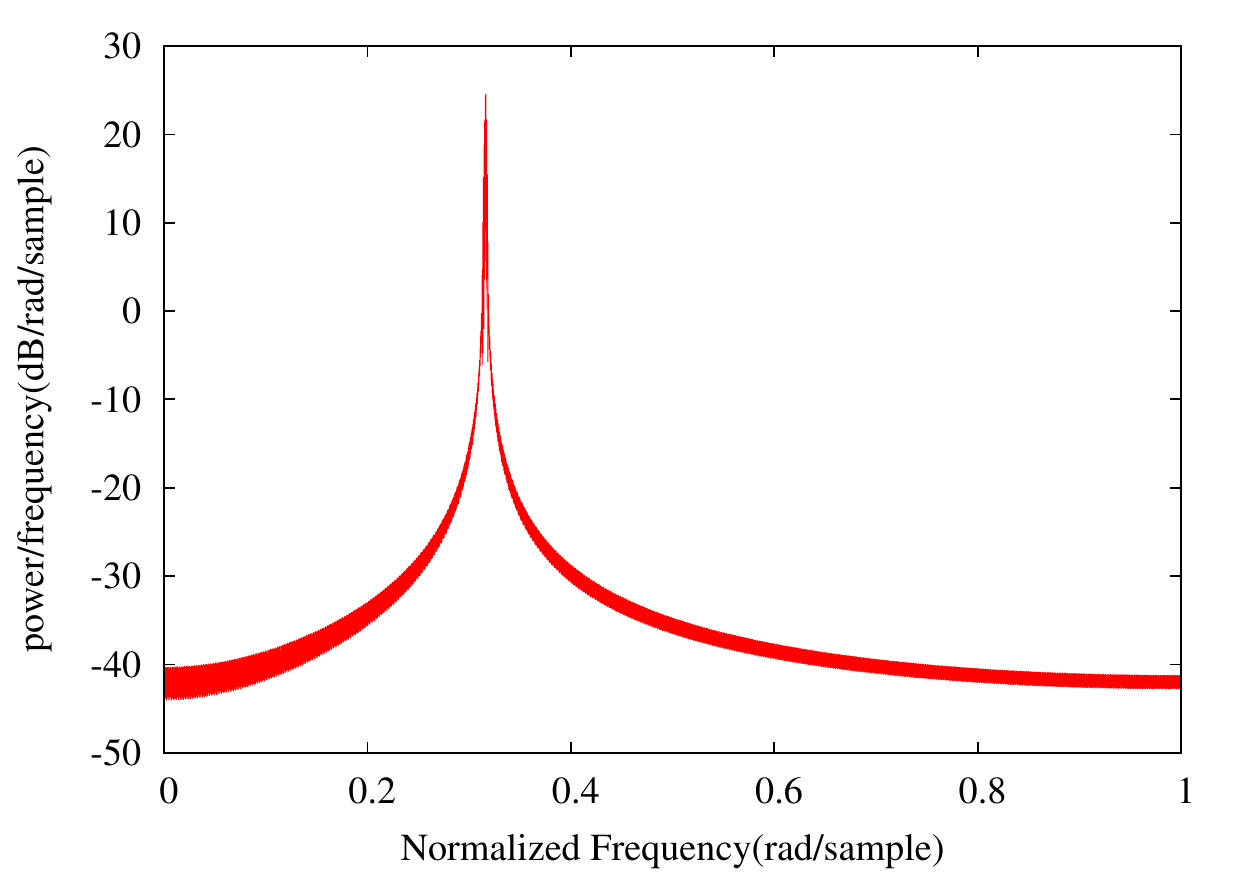}}
	\caption{Power spectra for $\alpha_{q}=2$ showing (a) chaotic nature ($q=0.9$), and (b) periodic nature ($q=0.999$).} 
	\label{powerspectrum}
\end{figure}

The power spectrum for the system, in the range of $0.2<q<0.99$ for $\alpha_q=1$, shows grassiness and decreasing trend, which is a typical feature of a chaotic power spectrum. An example of such a spectrum is given in Fig. \ref{powerspectrum}(a). The decreasing trend is in such a way that initially there is an exponential decay which is followed by a much slower algebraic decay. The grassiness is more if $\alpha_{q}=2$ than if $\alpha_{q}=1$. One expects multiple (and/or split) peaks in a quasi-periodic power spectrum. As such, we observe, two distinct split peaks near zero for $q=0.2$ when $\alpha_q=1$, thus re-affirming our observations regarding its quasi-periodic behaviour. Our analysis also shows quasi-periodic behaviour in the range $0.1<q\leq0.2$ when $\alpha_q=1$. The resurgence of an exact periodic behaviour is observed when $q\leq0.1$ for $\alpha=1$. For the quasi periodic behaviour arising when $\alpha_{q}\leq0.5$ at $q=0.95$, we observe two peaks, affirming the presence of two distinct periods.

Based on the qualitative understanding of the system from the previous sections, we now move onto the quantitative analysis of the system.
\subsection{First-return-time distributions}  
\begin{figure}[h!]
	\centering
	\subfloat[$q$=0.1]{\includegraphics[width=0.5\linewidth]{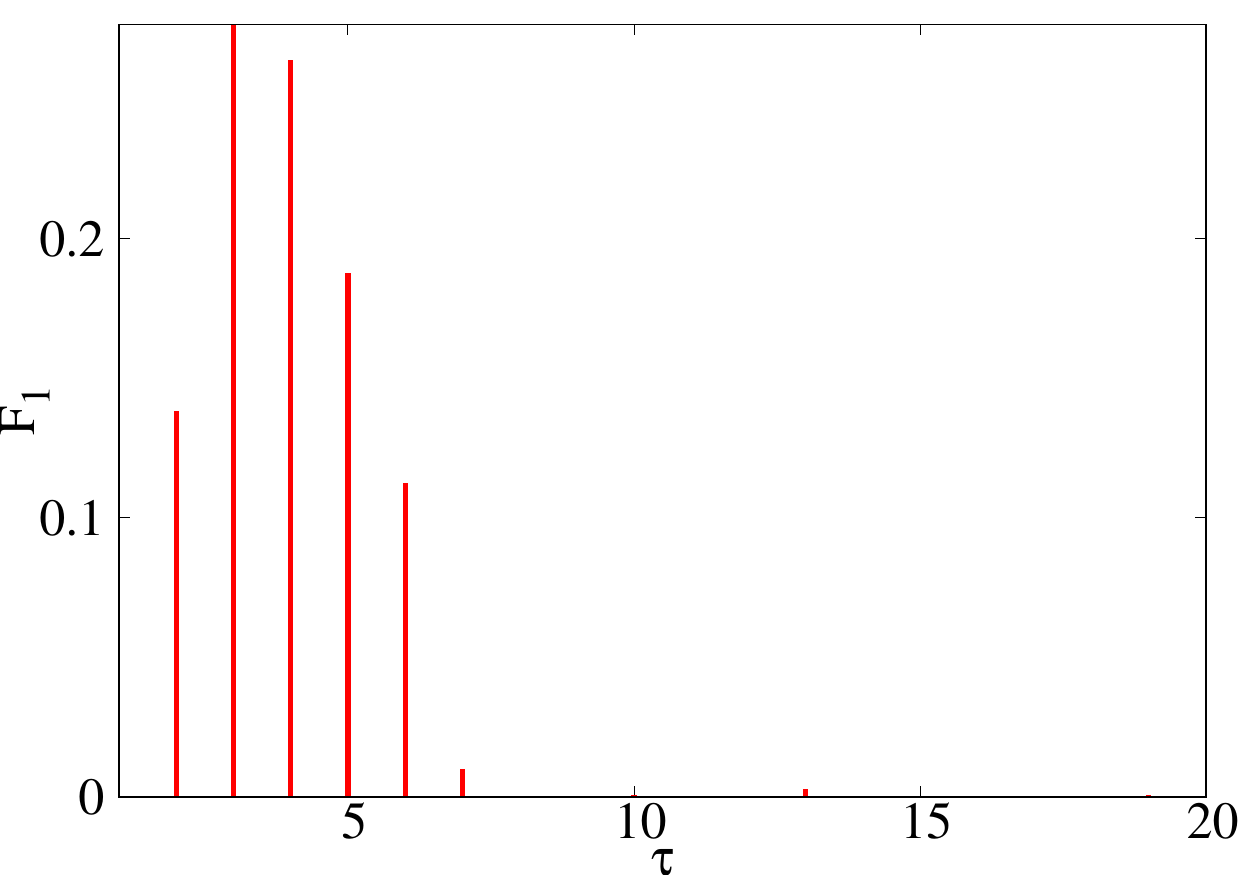}}
	\subfloat[$q$=0.2]{\includegraphics[width=0.5\linewidth]{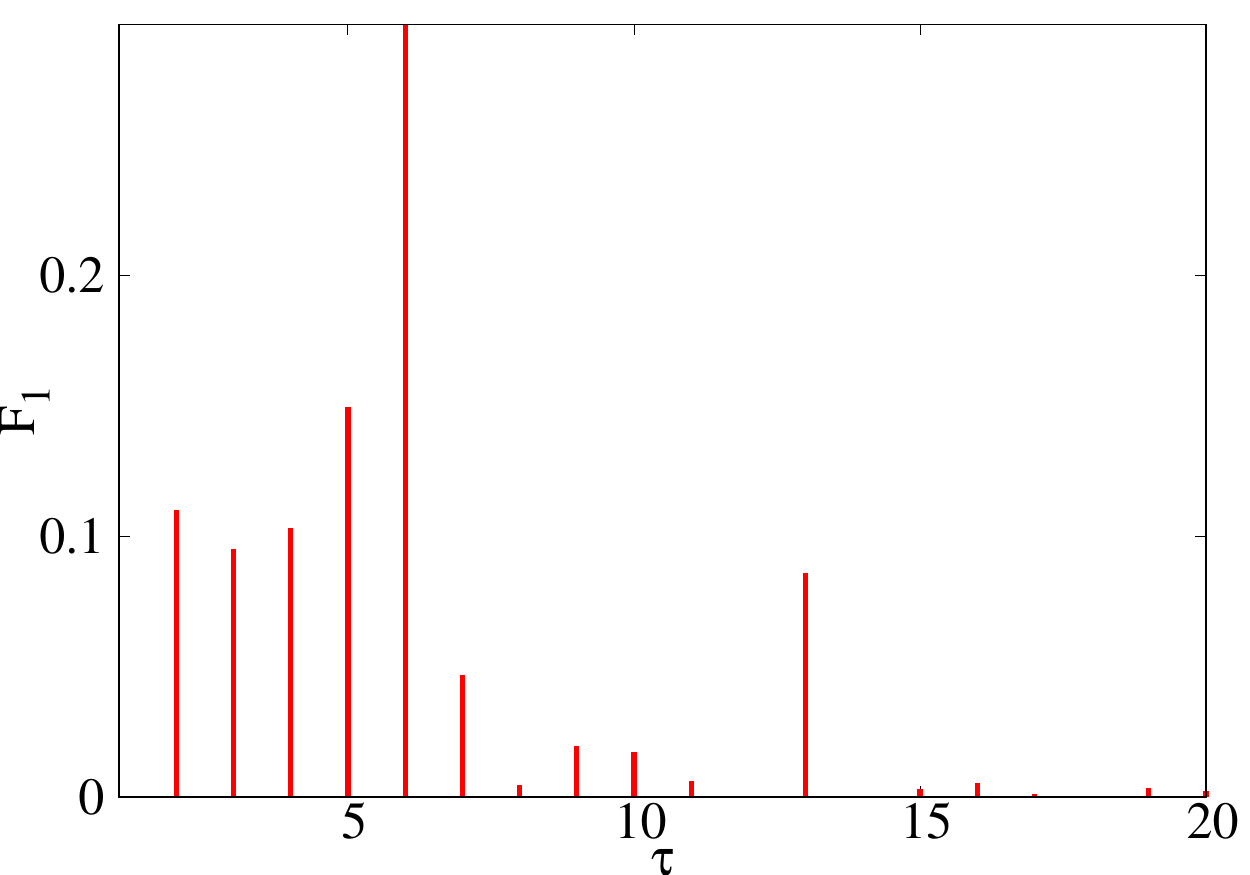}}\\
	\subfloat[$q$=0.9]{\includegraphics[width=0.5\linewidth]{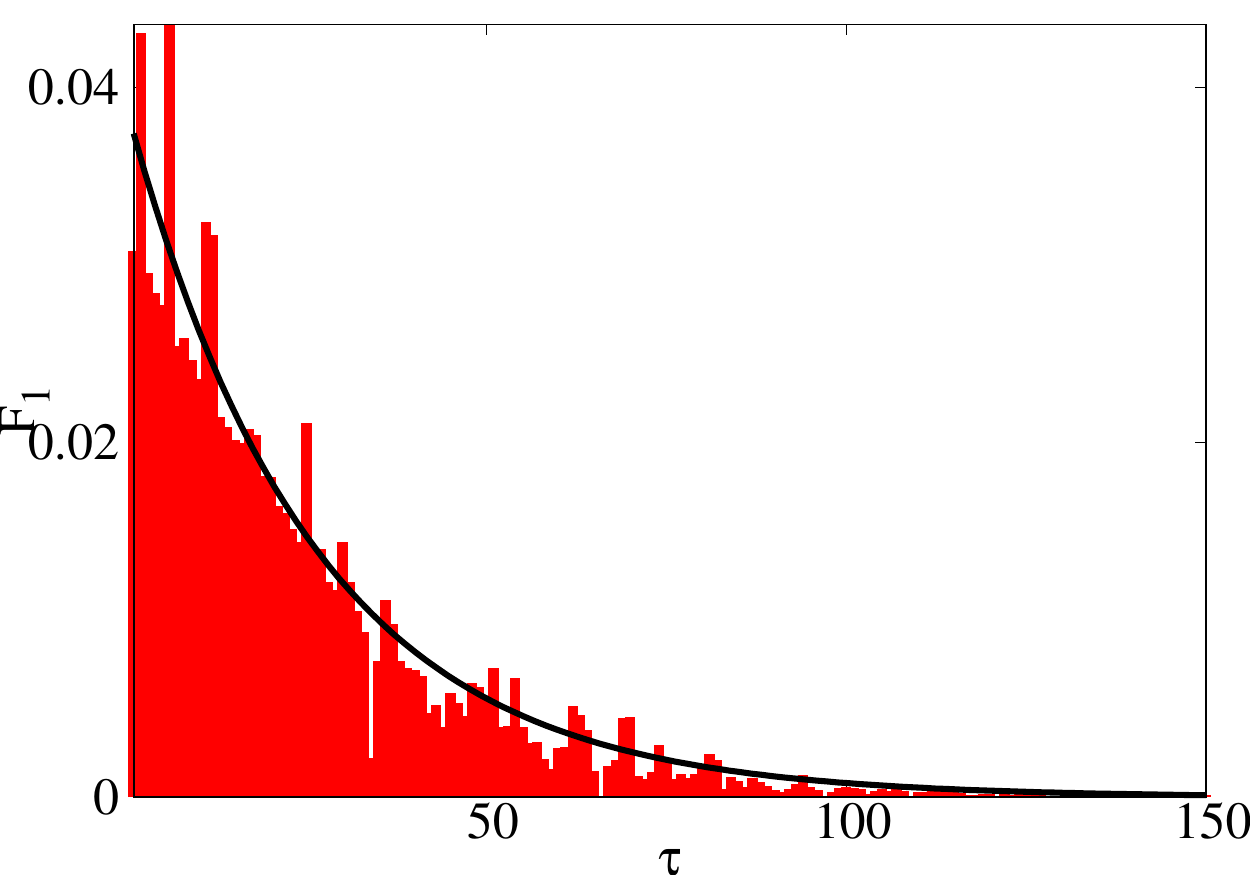}}
	\subfloat[$q$=0.99]{\includegraphics[width=0.5\linewidth]{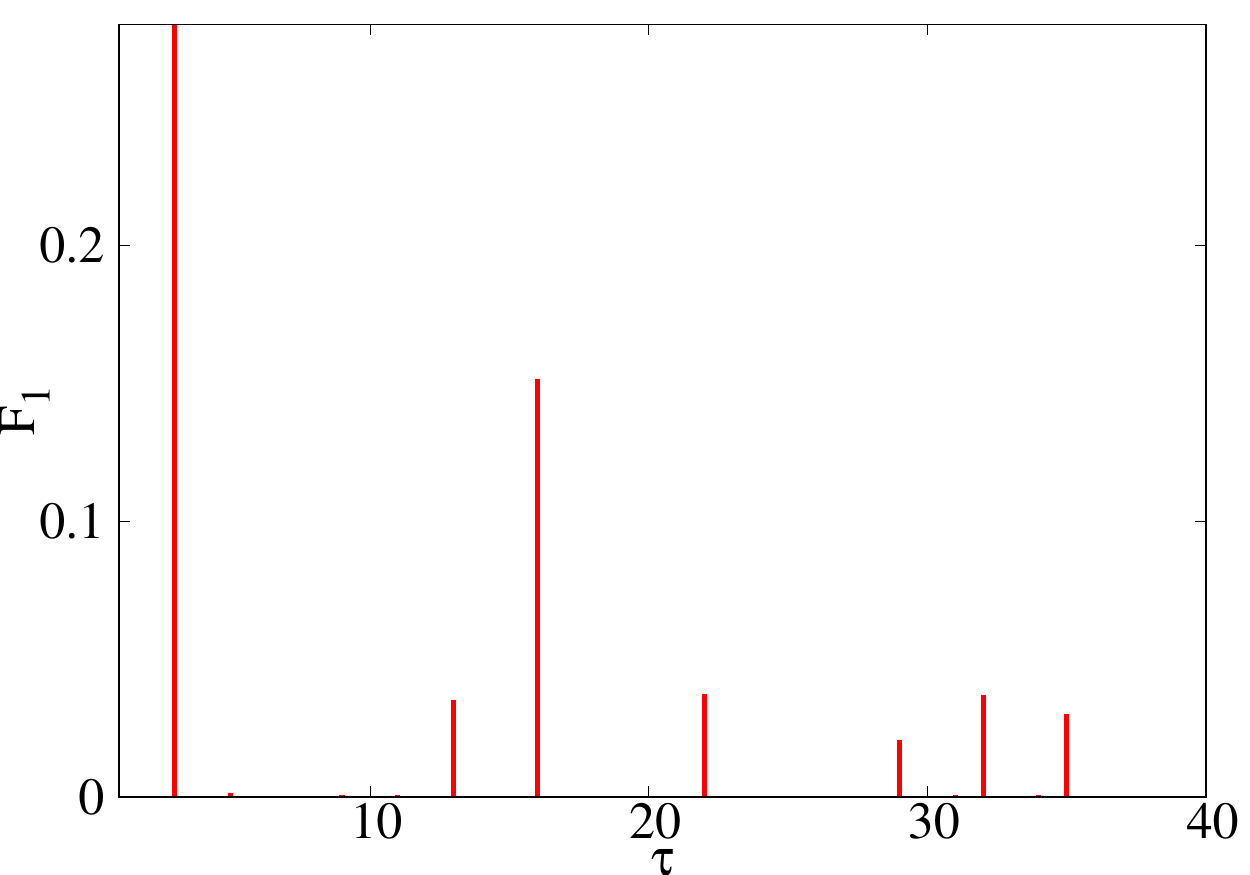}}
	\caption{First-return-time distributions for different values of $q$ and $\alpha_{q}=1$. (a) corresponds to periodic data, (b), (d) corresponds to quasi-periodic data and (c) corresponds to chaotic data. The solid black line in the first-return-plot represents the fitted exponential curve mentioned in section \ref{sub3}.}
	\label{fr_j1}
\end{figure}

 Figure \ref{fr_j1} presents the first-return-time distributions. The first-return-time plots for those time series that were observed to produce chaotic behaviour earlier are seen to fit the exponential probability density function explained in \cite{Sudheesh2}. This points to their ergodic dynamical behaviour. When $\alpha_{q}=1$, traces of chaotic behaviour is observed around $q=0.3$ and becomes more evident as $q$ increases. Around this initial chaotic region, although the distribution resembles an exponential decay, it fits only approximately to the said distribution. The expectation values for $\alpha_q=1$ in the range $0.1<q\leq0.2$ seems to exhibit a quasi-periodic behaviour, characterised by the distribution of the type shown in Fig. \ref{fr_j1}(b). In the case of exponential fitting, the mean recurrence time, $\mu$, is found to be large, thus indicating large time variation before recurrence. For $q\leq0.1$, the behaviour strictly adheres to periodicity, thus not fitting any special distributions.

Similarly, for low $\alpha_{q}$ featuring a band-like phase space structure, the first-return-time analysis revealed quasi-periodic behaviour. Higher $\alpha_{q}$ values produce chaotic behaviour for a larger range of the allowed $q$ values, while lower $\alpha_{q}$ values show more quasi-periodic and periodic properties for a larger range of $q$ values. Comparing the dynamical behaviour for $\alpha_{q}=1$ and $\alpha_{q}=2$, we find that chaotic behaviour is prevalent for nearly all allowed $q$ values for $\alpha_q=2$ while it is seen only for $0.2< q<0.99$ when $\alpha_q=1$. For lower values of $\alpha_{q}$, the system is periodic for all values of $q$. 


\subsection{Lyapunov exponent of the time series}
\label{sub32}
The maximum Lyapunov exponents of the time series evaluated for different combinations of $\alpha_{q}$ and $q$ values were found to be positive for those ranges that were deduced to be chaotic in the previous section. The plot of the logarithm of distance between two closely spaced trajectories versus time is of particular importance. This curve is ideally expected to be a straight line, whose slope gives the magnitude and sign of the Lyapunov exponent. The plot for different $q$ values and different $\alpha_{q}$ values were determined and plotted, with the results in agreement with our findings in the previous sections. 
\begin{figure}[h!]
	\centering 
	\includegraphics[width=0.8\linewidth]{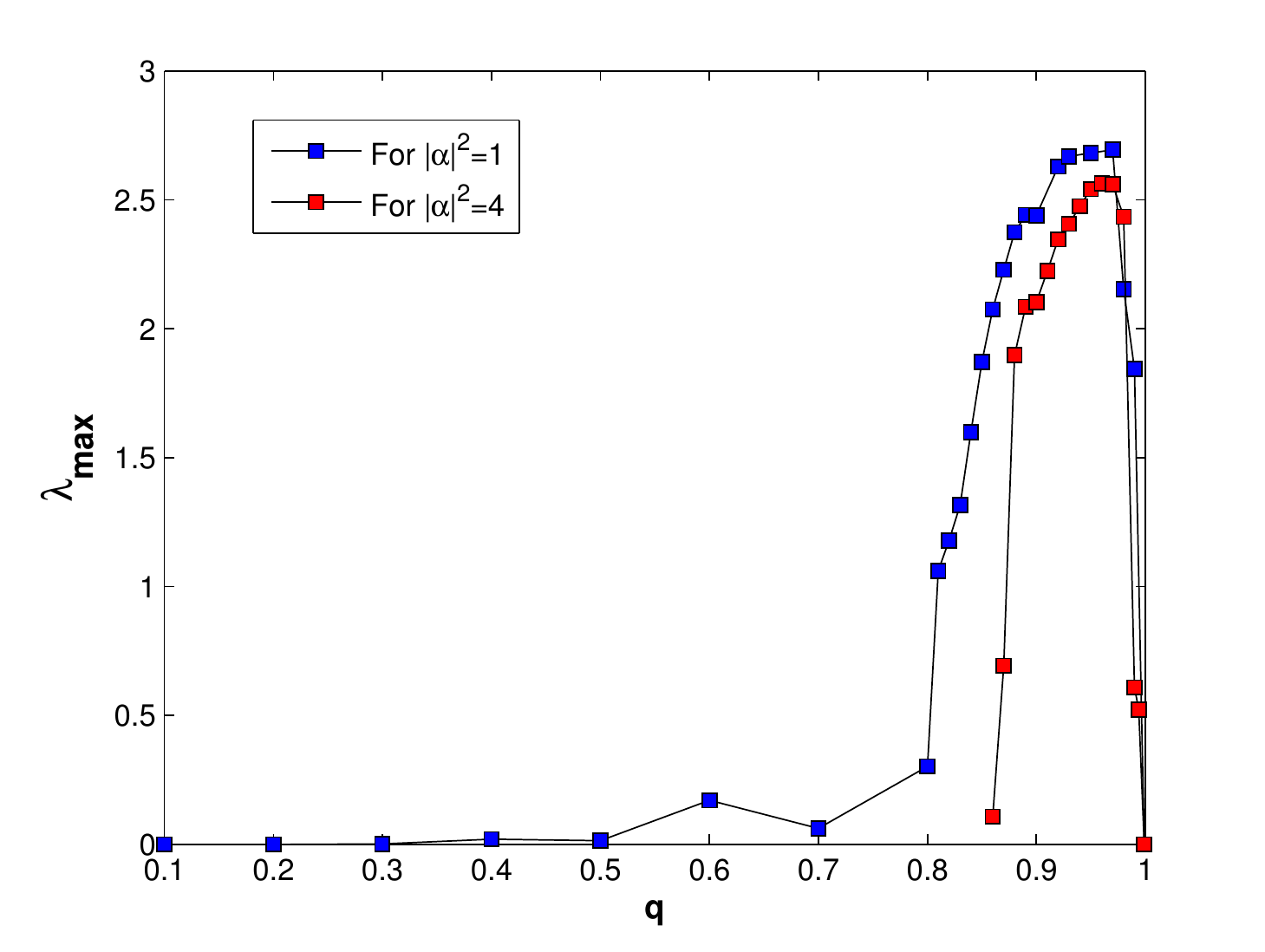}
	\caption{Variation of largest Lyapunov exponent with $q$ value.}
	\label{lyapunov}  
\end{figure}

\begin{figure}[h!]
	\centering 
	\includegraphics[width=0.8\linewidth]{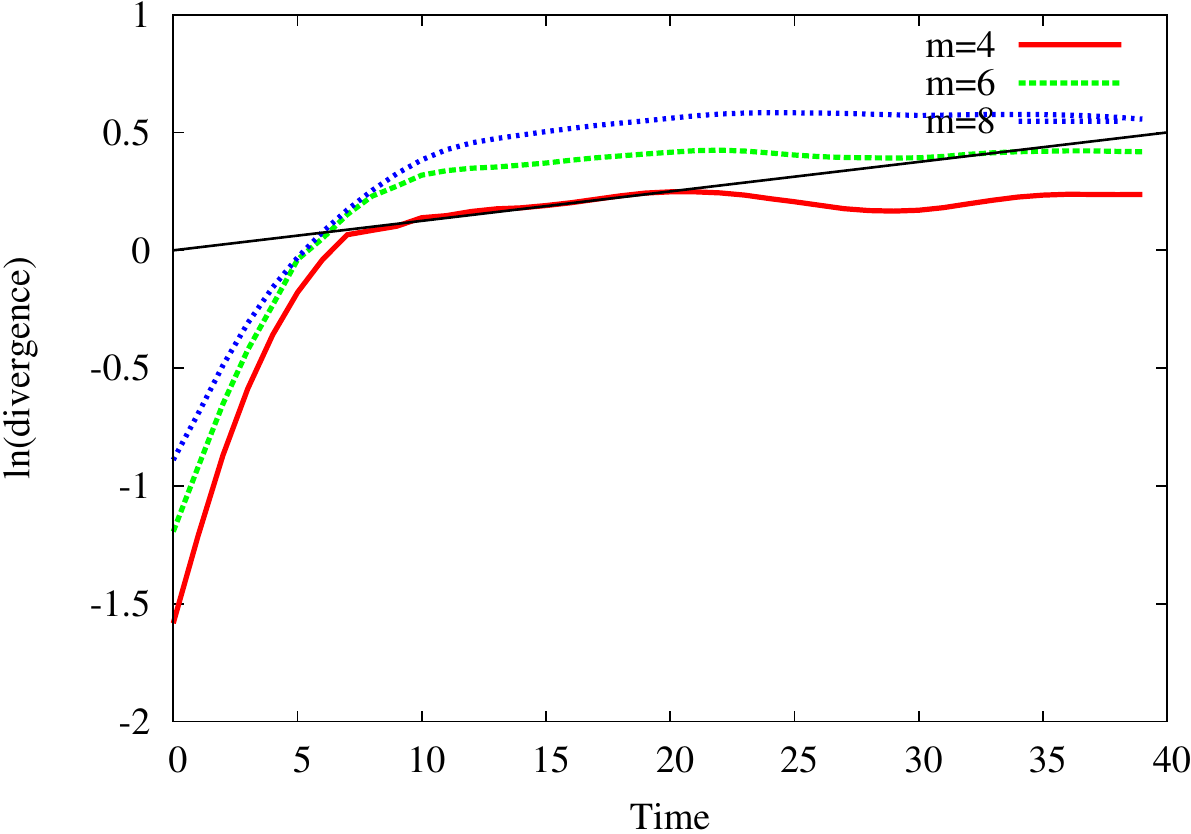}
	\caption{Lyapunov exponent plot for $\alpha_{q}=1$ and $q=0.9$ showing the chaotic nature of the system.}
	\label{lya2}  
\end{figure}

 The sign of the Lyapunov exponent is unambiguously confirmed to be positive for $0.2< q<0.99$ when $\alpha_q=1$, which is a signature of chaos. Further, the slope of the curve for the periodic and quasi-periodic regime is found to be $0$, thus re-affirming our conclusions regarding their behaviour as well. Fig. \ref{lyapunov} shows the variation of $\lambda_{max}$ with $q$. For a given value of $\alpha_{q}$, the transition to periodicity is faster in the region of $q\rightarrow 1$ than when $q\rightarrow 0$. We also include Fig. \ref{lya2} to portray the nature of the curves used to determine the largest Lyapunov exponents \cite{Rosenstein}. Here, the linear region is used for the calculation of the exponent.

\begin{figure}[h!]
	\centering 
	\includegraphics[width=1\linewidth]{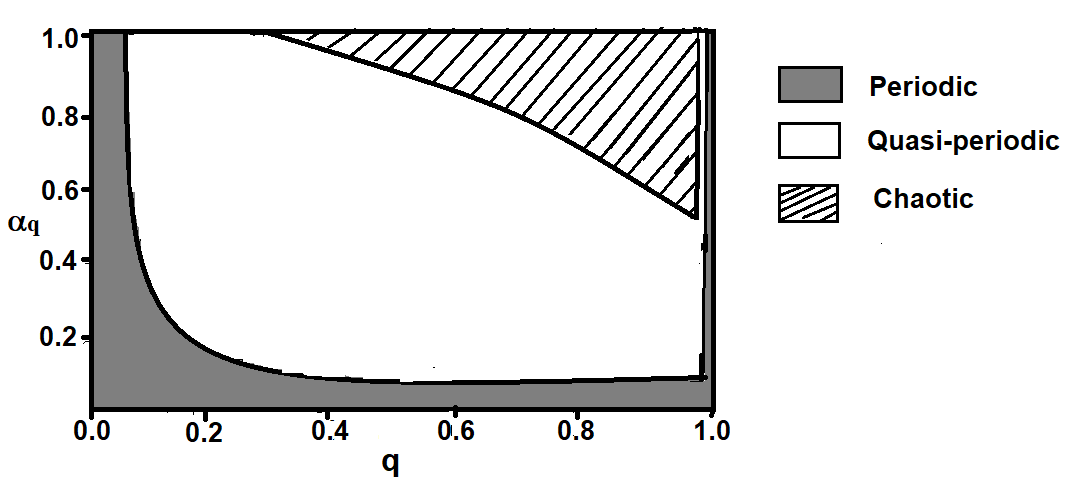}
	\caption{Behaviour of a $q$-deformed oscillator with respect to $\alpha_{q}$ and $q$ values.}
	\label{result}
\end{figure}

As we come to the conclusion, with all the proven results, we provide an approximate demarcation of the various dynamical regimes in the $q-\alpha_{q}$ plane using Fig. \ref{result}.\\

\section{Conclusion}
\label{sec4}
In this study, we attempted to understand how the $q$-deformation of an ordinary Hamiltonian changes the behaviour of the system. We find that the $q$-deformation confers non-linear properties to the ordinary quantum harmonic oscillator. By studying the dynamics of the resultant times series obtained for the expectation values of the dynamical variables $X$ and $P$, we conclude that the system studied exhibits periodic, quasi-periodic and chaotic behaviour depending on the deformation parameter $q$ and the deformed coherent amplitude $\alpha_{q}$. As the value of $\alpha_{q}$ increases, we observe that chaotic nature is the more prominent dynamical behaviour. For lower values of $\alpha_{q}$, the dynamical behaviour is mostly quasi-periodic or periodic.

The qualitative verification of the dynamical properties of this system was performed through recurrence plots and power spectra of the time series. First-return-time distributions were used to verify the ergodic behaviour of the system in the chaotic regimes obtained from the qualitative analysis. The quantitative verification of the above conclusions using Lyapunov exponents revealed that the exponents are positive in the estimated chaotic regime, and zero in the periodic and quasi-periodic regimes. The magnitude of the Lyapunov exponents and thus the magnitude of the exponential divergence of trajectories is dependant on the magnitude of the deformation. 

Thus, the analysis of expectation values of dynamical variables carried out in this paper clearly shows signatures of chaos in another quantum system.

\end{document}